\documentclass[%
 reprint,
 showkeys,
superscriptaddress,
 amsmath,amssymb,
]{revtex4-1}
\usepackage{graphicx}
\usepackage{dcolumn}
\usepackage{bm}
\usepackage{siunitx}\usepackage{bm}
\usepackage{siunitx}
\usepackage{color,soul}

\usepackage[utf8]{inputenc}
\usepackage{caption}
\begin{document}
\title{Phenomenological classification of metals based on resistivity}
\author{Qikai Guo}
\email{Corresponding author: qikaiguo@sdu.edu.cn}
\affiliation{Zernike Institute for Advanced Materials, University of Groningen, The Netherlands}
\affiliation{School of Microelectronics, Shandong University, Jinan, 250100, China}
\author{César Magén} 
\affiliation{Instituto de Nanociencia y Materiales de Aragón (INMA), CSIC-Universidad de Zaragoza, 50009 Zaragoza, Spain}
\affiliation{Laboratorio de Microscopías Avanzadas (LMA), Universidad de Zaragoza, 50018 Zaragoza, Spain}
\author{Marcelo J. Rozenberg}
\affiliation{Université Paris-Saclay, CNRS, Laboratoire de Physique des Solides, 91405 Orsay, France}
\author{Beatriz Noheda}
\email{Corresponding author: b.noheda@rug.nl}
\affiliation{Zernike Institute for Advanced Materials, University of Groningen, The Netherlands}
\affiliation{CogniGron center, University of Groningen, The Netherlands}

\begin{abstract}

Efforts to understand metallic behaviour have led to important concepts such as those of strange metal, bad metal or Planckian metal. However, a unified description of metallic resistivity is still missing. An empirical analysis of a large variety of metals shows that the parallel resistor formalism used in the cuprates, which includes $T$-linear and $T$-quadratic dependence of the electron scattering rates, can be used to provide a phenomenological description of the electrical resistivity in all metals, where these two contributions are shown to correspond to the two first terms of a Taylor expansion of the resistivity, detached of their physics origin, and thus, valid for any metal. Here we show that the different metallic classes are then determined by the relative magnitude of these two components and the magnitude of the extrapolated residual resistivity. These two parameters allow to categorize a few systems that are notoriously hard to ascribe to one of the currently accepted metallic classes. This approach also reveals that the $T$-linear term has a common origin in all cases, strengthening the arguments that propose the universal character of the Planckian dissipation bound.

\end{abstract}

\maketitle

\section{INTRODUCTION}

Interactions of electrons with other (quasi-)particles (e.g. phonons, magnons or electrons themselves) are responsible for the electrical transport of metallic systems. In simple metals, electron-electron interactions lead to a Fermi liquid description \cite{landau1959theory} of the resistivity at low temperatures ($T$) as a $T^2$ dependence; while a linear increase of resistivity is usually observed at high $T$ because of the boosted scattering strength between electrons and phonons. However, this well-defined regime meets with problems in strongly correlated metals. This can happen when the metallic system is driven close to a Quantum Critical Point, which gives rise to a $T$-dependence of resistivity of the type $T$\textsuperscript{n}, with 1$\leq$ n $<$ 2 at low-$T$ \cite{Stewart2001,lee2018recent}. Among those, the "strange metals", such as the optimally-doped cuprates, present a puzzling linear-$T$ dependence of resistivity that can range from very low $T$ (thus discarding phonon scattering) to high $T$.

Deviations from the standard behavior also take place at high temperatures, when the increased scattering drives the mean-free path ($\ell$) to approach the Mott-Ioffe-Regel (MIR) limit \cite{ioffe1960non}\footnote{The Mott-Ioffe-Regel (MIR) limit reflects the electrons mean-free-path approaching the interatomic spacing.}, which compels the resistivity to show saturation at high-$T$ \cite{gunnarsson2003colloquium}. However, in some \textit{so-called} incoherent or "bad metals", the resistivity overcomes this upper limit, implying such large scattering rates that, according to Heisenberg’s principle, the uncertainty in the quasi-particles energy prevents their coherence, thus, disqualifying the quasi-particle description altogether \cite{hussey2004universality,takenaka2002coherent}.

Different conduction mechanisms become dominant at different temperatures and, thus, an overall description of metallic resistivity over a wide temperature range requires considering the combined effect of the various contributions in a phenomenological manner. $T$-linear resistivity has been typically associated to electron-phonon scattering and, thus, such dependence was not expected at low temperatures. However, it is now well established that $T$-linear resistivity can emerge well below the Debye temperature in various systems, from simple metals to strongly correlated metals, as long as the scattering rate (1/$\tau$) per Kelvin of the charge carriers reaches a universal bound, $k$\textsubscript{B}/$\hbar$. This so-called "\textit{Planckian"} dissipation limit (PDL) is independent of the distinct behaviour and conduction mechanism \cite{bruin2013similarity}. These findings challenge the significance of a specific scattering mechanism in determining "strange" metallic transport and motivates the search for a phenomenological description that applies to a large variety of metals. In this paper we propose such a description. 

\section{RESULTS AND DISCUSSION}

\begin{figure*}
\centering
\includegraphics[width=1\textwidth]{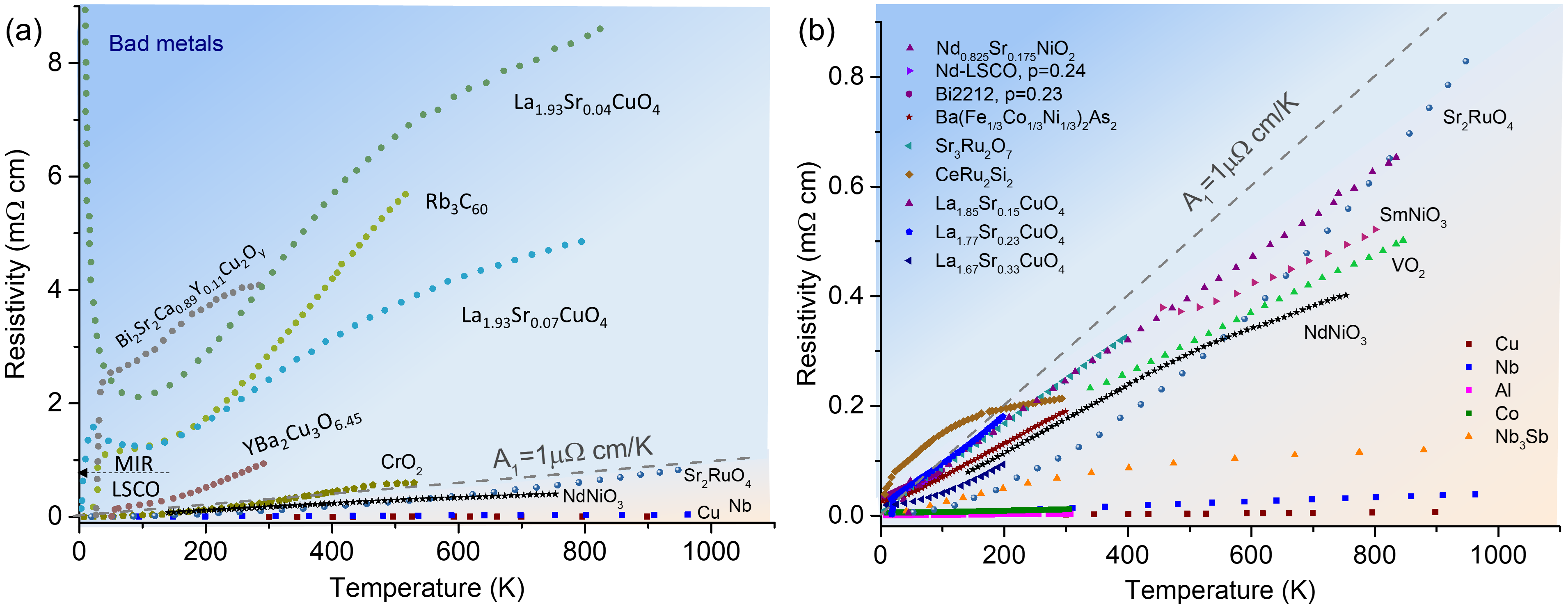}
\caption{Resistivity of various metallic systems. (a) Compilation plot showing the resistivity of various metallic systems. (b) Blow-up of the intermediate region in (a). The dashed line indicates the linear-$T$-resistivity slope of 1 $\mu$ $\Omega$ cm; the arrow shows the MIR limit (0.7 m$\Omega$ cm) of La\textsubscript{2-x}Sr\textsubscript{x}CuO\textsubscript{4}. (Data source: Cu, Nb \cite{gunnarsson2003colloquium}; Al, Co, and Pd \cite{de1988temperature}; La\textsubscript{2-x}Sr\textsubscript{x}CuO\textsubscript{4} with x=0.17-0.33 \cite{cooper2009anomalous}, with x=0.04 and 0.07 \cite{takagi1992systematic}; Bi\textsubscript{2}Sr\textsubscript{2}Ca\textsubscript{0.89}Y\textsubscript{0.11}Cu\textsubscript{2}O\textsubscript{y} \cite{wang1996observation}; Rb\textsubscript{3}C\textsubscript{60} \cite{hebard1993absence}; YBa\textsubscript{2}Cu\textsubscript{3}O\textsubscript{6.45} \cite{ito1993systematic}; Sr\textsubscript{2}RuO\textsubscript{4}  \cite{tyler1998high}; CeRu\textsubscript{2}Si\textsubscript{2} \cite{besnus1985low}; CrO\textsubscript{2} \cite{lewis1997band}; VO\textsubscript{2} \cite{allen1993resistivity}; SmNiO\textsubscript{3} \cite{jaramillo2014origins}; Nd\textsubscript{0.825}Sr\textsubscript{0.175}NiO\textsubscript{2}  \cite{li2020superconducting}; Nd-LSCO (p=0.24) and Bi2212 (p=0.23) \cite{legros2019universal}; Ba(Fe\textsubscript{1/3}Co\textsubscript{1/3}Ni\textsubscript{1/3})\textsubscript{2}As\textsubscript{2} \cite{nakajima2020quantum}; Sr\textsubscript{3}Ru\textsubscript{2}O\textsubscript{7} \cite{bruin2013similarity}.) }
\label{fig:summary of metallic systems}
\end{figure*}

In Fig. \ref{fig:summary of metallic systems}(a), the $\rho$(T) curves of a wide diversity of metallic systems are plotted together. These systems include cuprates with different doping levels, ruthenates, heavy fermions, alkali-doped $C$\textsubscript{60}, iron pnictides, transition metals and monovalent metals. These materials have been classified as simple metals, correlated metals, strange metals, bad metals or Planckian metals. Compared with the slowly increasing resistivity of simple metals (in the yellow region), for the metals displayed in the blue region, the slope of $\rho$(T) at around 300 K (which in most of cases is the maximum slope) is large, as expected from strong electron scattering. Among them, the under-doped cuprates and Rb\textsubscript{3}C\textsubscript{60} are well-established bad metals \cite{gunnarsson2003colloquium}. The resistivity in these systems can cross the $\rho$\textsubscript{MIR} limit at relatively low temperatures and approach a value far beyond it at high temperatures, violating the quasi-particle scenario. 

Notably, most correlated metals are located in an intermediate region between the good metals and the bad metals, as shown with more detail in the blow-up plot of Fig. \ref{fig:summary of metallic systems}(b). Interestingly, a number of these intermediate systems remain unclassified, such as Sr\textsubscript{2}RuO\textsubscript{4} \cite{tyler1998high} and  CrO\textsubscript{2} \cite{lewis1997band}, which have been reported to possess properties of both conventional and bad metals. Other systems, like Sr\textsubscript{3}Ru\textsubscript{2}O\textsubscript{7}, Nd-LSCO (p=0.24), Bi2212 (p=0.23), and Ba(Fe\textsubscript{1/3}Co\textsubscript{1/3}Ni\textsubscript{1/3})\textsubscript{2}As\textsubscript{2}, have been discussed as Planckian metals \cite{bruin2013similarity, legros2019universal, nakajima2020quantum}. However, regardless of their different origin and classification, the resistivity of all the metallic systems in the intermediate region show many comparable features. For instance, the maximum slope of resistivity in most of materials is well below an upper limit of 1 $\mu \Omega$ cm/K. The same limit has been reported in high-$T$\textsubscript{c} cuprates and was associated with the momentum-averaged scattering rate ($\hbar /\tau$ $\sim$ $\pi k$\textsubscript{B}T) \cite{hussey2011dichotomy}, which corresponds to the PDL. As mentioned earlier, the PDL concept has been put forward as the common origin of the linear-$T$-resistivity in systems with very different scattering mechanisms, including  high-$T$\textsubscript{c} superconductors, other electron-correlated systems and even simple metals \cite{bruin2013similarity}.

In an effort to unify the behaviour of the different metallic systems, we follow the seminal work of Hussey \textit{et al.} in high-$T$\textsubscript{c} cuprates \cite{cooper2009anomalous}, which shows that the resistivity of La\textsubscript{2-x}Sr\textsubscript{x}CuO\textsubscript{4} at various doping levels can be successfully described by a parallel resistor formalism \cite{wiesmann1977simple} as:

\begin{equation} 
\label{eq for pallaror resistivity}
   \rho(T)^{-1}=\rho \textsubscript{ideal}(T)^{-1}+\rho \textsubscript{sat}^{-1}
\end{equation}
where $\rho$\textsubscript{ideal} is the resistivity in the absence of saturation, which is shunted by the large value of $\rho$\textsubscript{sat} at high temperatures. An adequate definition of $\rho$\textsubscript{ideal} is then needed in order to describe $\rho$(T) in a wide temperature range. Based on a large body of experimental data, dual-component model, with linear and quadratic terms, has been used in the cuprates \cite{hussey2008phenomenology,cooper2009anomalous} as:

\begin{equation}\label{definiation of ideal resistivity}
\rho \textsubscript{ideal}(T)=\rho_{0}+A_1 T + A_2 T^2
\end{equation}
where $\rho_{0}$ represents the residual resistivity, and the other two terms reflect the temperature dependence of the scattering rate with an isotropic $T$-quadratic component (assigned to electron–electron scattering) and an anisotropic $T$-linear component that is consistent with the PDL \cite{cooper2009anomalous}. Generalizing this to other types of metals, $A_1$ and $A_2$ cannot be assigned to a specific scattering mechanism and the previous equation should be generally considered as a Taylor expansion of $\rho$ (T).


\begin{figure*}
\centering
\includegraphics[width=0.7\textwidth]{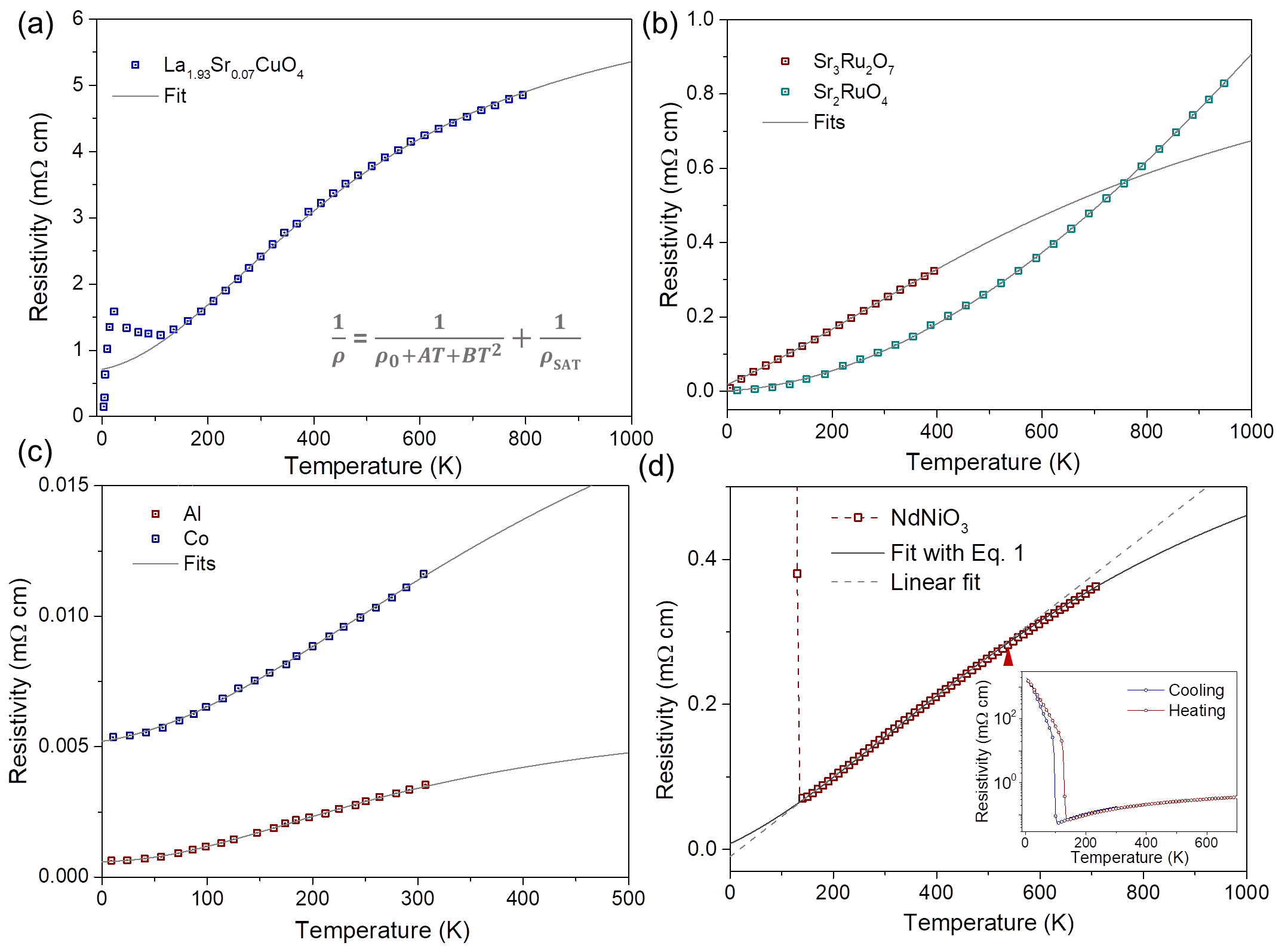}
\caption{Fitting of the electrical resistivity of various metallic systems using Eq. \ref{eq for pallaror resistivity}. (a) Underdoped cuprate; (b) Ruthenates; (c) Simple metals; (d) Nickelate (NdNiO$_3$). The triangle in (d) indicates the temperature above which the data deviates from a $T$-linear dependence. (Data sources: resistivity of NdNiO\textsubscript{3} is measured in the present work; while data of other systems are extracted from Refs. \cite{takagi1992systematic,tyler1998high,bruin2013similarity,de1988temperature}, respectively.) }
\label{fig:fit to other systems}
\end{figure*}

 \begin{figure*}
\centering
\includegraphics[width=0.95\textwidth]{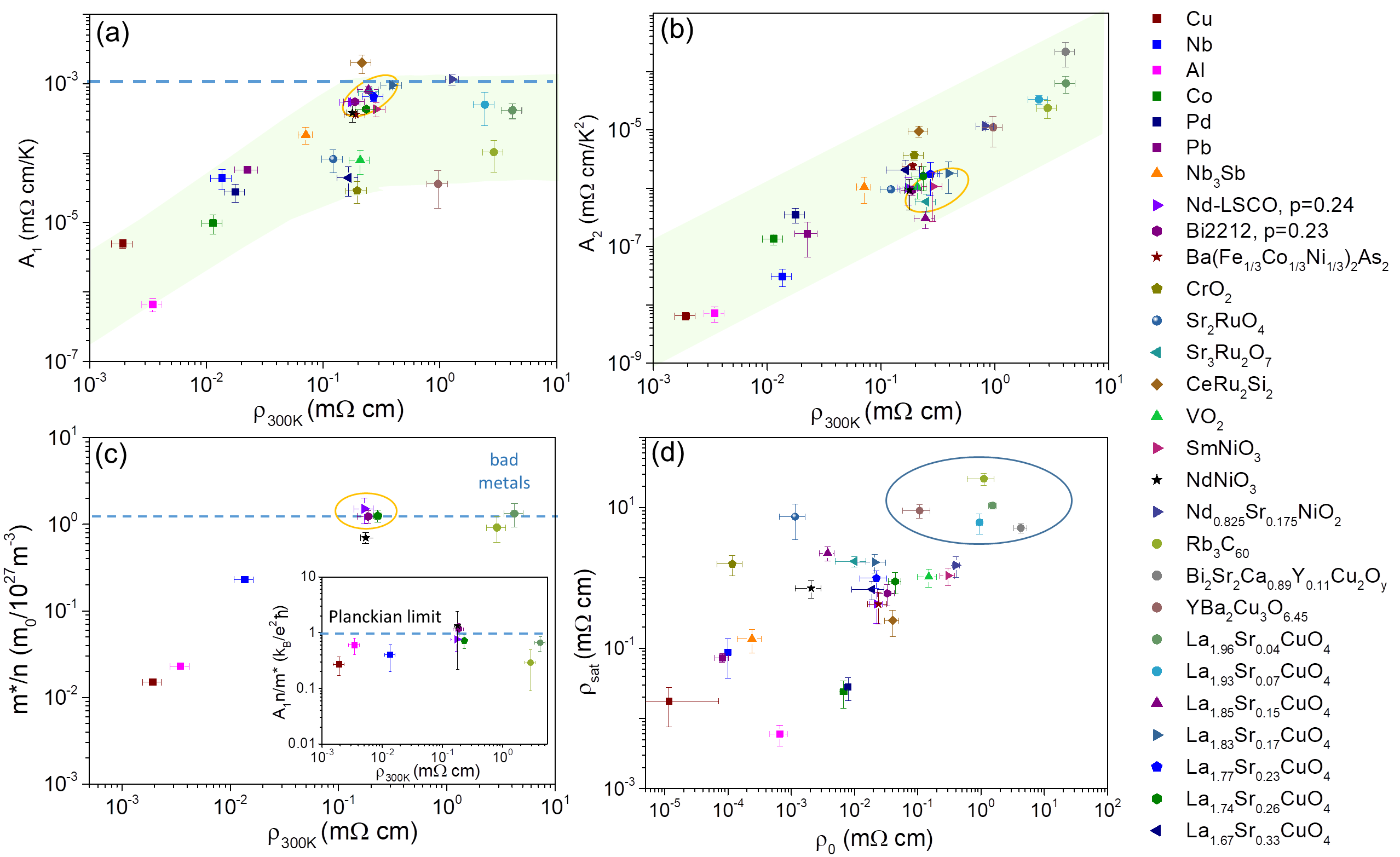}
\caption{Observed trends in the DC-PRF coefficients. From the fit of $\rho$(T) in various metals to the DC-PRF model of Eqs. \ref{eq for pallaror resistivity}-\ref{definiation of ideal resistivity}, the coefficients (a) $A$\textsubscript{1} and (b) $A$\textsubscript{2} are obtained and plotted as a function of the corresponding room temperature resistivity ($\rho$\textsubscript{300K}). The blue dashed line in (a) indicates the maximum value obtained for $A$\textsubscript{1} $\sim$ 1 $\mu \Omega$ cm/K. Error bars are also obtained from the fitting results. Encircled in yellow are the strange metals. Shadows are a guide-to-the-eye showing the general evolution of the coefficients. (c) the m*/n ratio is plotted as a function of $\rho$\textsubscript{300K}, showing a trend similar to $A$\textsubscript{1}. In the inset, $A$\textsubscript{1}${n/m*}$ is shown to remain at, or slightly below, $k$\textsubscript{B}/($e$\textsuperscript{2}$\hbar$), which corresponds the Planckian dissipation limit (PDL), also for the normal metals and bad metals. (d) Saturation resistivity ($\rho$\textsubscript{sat}) as a function  of residual resistivity ($\rho$\textsubscript{0}). Encircled in blue are the bad metals.}
\label{fig: extracted parameters}
\end{figure*}


Here we show that this dual-component parallel-resistor formalism (DC-PRF) can describe the metallic behaviour of very distinct systems, independently of the dominant scattering mechanism. The DC-PRF has been used to fit the remarkable variety of resistivity data shown in Fig. 1, from the bad metals to the good metals. As shown in Fig. \ref{fig:fit to other systems}(a)-(c) and Supplementary Material, the electrical resistivity of all these materials can be well described by Eqs. \ref{eq for pallaror resistivity}-\ref{definiation of ideal resistivity}. In all cases  $A_2 <  A_1$, reflecting a strong linear component at low temperatures (see Fig. S34).  This analysis provides us with a unified view of metallic behaviour. As shown in Fig. \ref{fig: extracted parameters} (a) and (b), the fitting of resistivity to all those metallic systems, reveals a clear evolution of $A$\textsubscript{1} and $A$\textsubscript{2} as a function of their room temperature resistivity ($\rho$\textsubscript{300K}). 
Interestingly, the data includes NdNiO$_3$ (NNO), which we consider in the present work as both, a test case and an illustration of the utility of the formalism. In fact, this compound is attracting significant attention, since superconductivity was reported in the related infinite layer system Nd$_{1-x}$Sr$_x$NiO$_2$ \cite{li2019superconductivity}, and is a remarkable example of a material whose metallic behavior has been particularly difficult to classify \cite{jaramillo2014origins,mikheev2015tuning,liu2013heterointerface}. For the present study, we have used epitaxial NNO films grown on LaAlO\textsubscript{3} substrates, which have been characterized in detail in our previous work \cite{guo2020tunable,guo2021hidden}. High-angle annular dark field (HAADF) STEM image shown in Supplementary Material Section 1 demonstrates the high crystalline quality of the NNO films with an atomically sharp interface with the substrate. In this 10 nm film, a first-order metal-to-insulator transition happens below 150 K. Here, measurements of resistivity in an extended temperature range allow for a clear determination of the $T$-dependence in the metallic state.  

 \begin{figure*}
\centering
\includegraphics[width=0.6\textwidth]{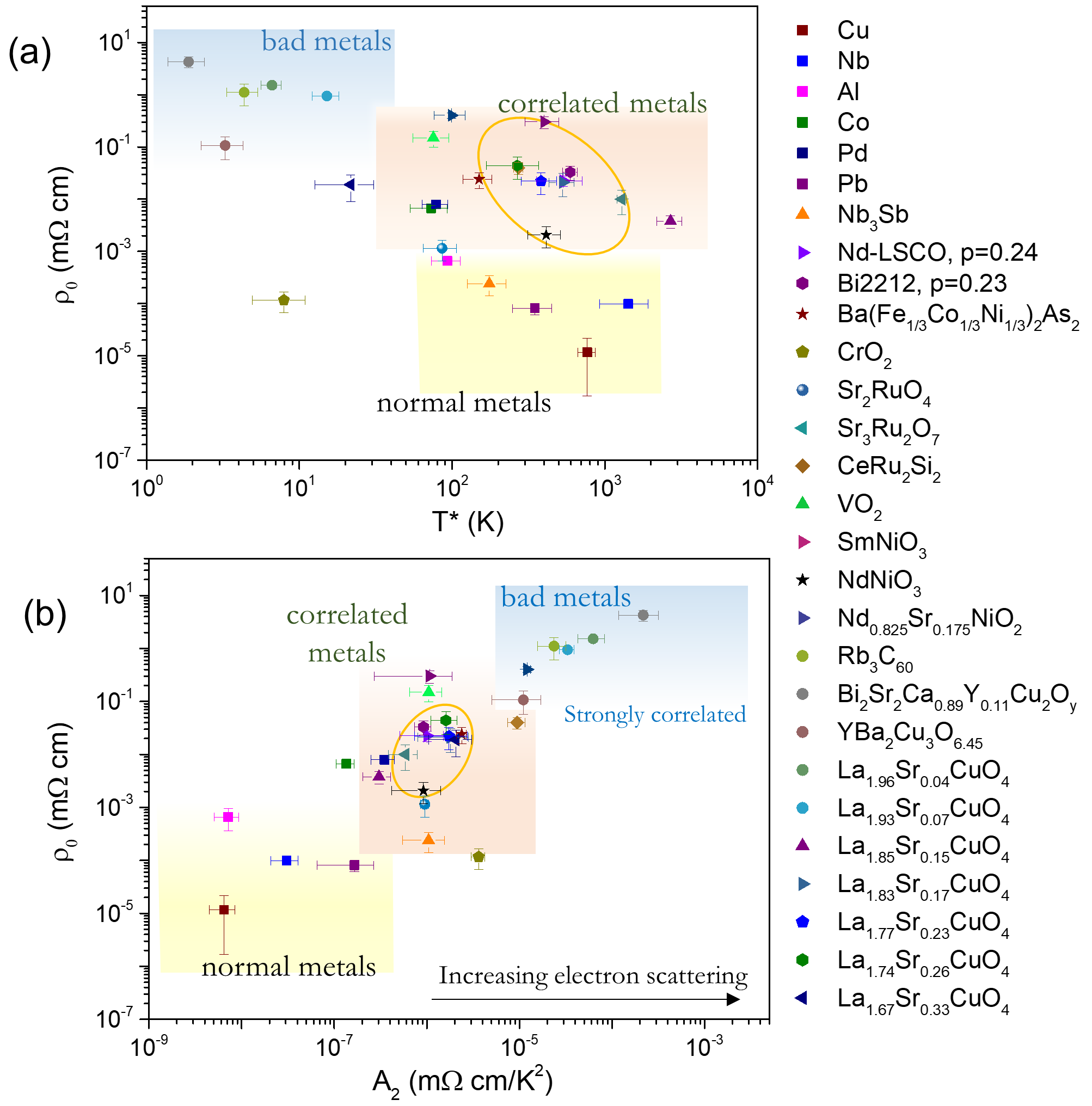}
\caption{Key parameters for the general classification of metals. (a) Residual resistivity ($\rho$\textsubscript{0}) \textit{versus} $T$*= $A$\textsubscript{1}/$A$\textsubscript{2}.  Bad metals, recognized by their large $\rho$\textsubscript{0}, are shown to display the largest $A$\textsubscript{2} and the lowest $T$* (the quadratic term dominates the scattering in the widest temperature range); while in strange metals (encircled in orange in all the figures) is the linear term the one dominating at most temperatures. (b)  $\rho$\textsubscript{0} versus $A$\textsubscript{2}, representing the magnitude of coherent contribution to the electron scattering rate.}
\label{fig: residual resistivity}
\end{figure*}

As shown in Fig. \ref{fig:fit to other systems}(d), a linear-$T$-resistivity is observed in NNO in a ultra-wide temperature range (about 400 K). In our previous works, we showed that this $T$-linear behaviour can be achieved in optimized NNO films with low epitaxial strain and low defect content \cite{guo2020tunable} and, more interestingly, it has signatures of Planckian dissipation \cite{guo2021hidden}. With the further increase of \textit{T}, the rise of $\rho$(\textit{T}) shows an obvious deviation from the linear dependence, which is caused by the addition of a parallel saturation resistance that takes over the behaviour in the high temperature regime \cite{wiesmann1977simple,gunnarsson2003colloquium,hussey2004universality}. Moreover, as in NdNiO\textsubscript{3} strong electron-electron interactions are expected, the combined effect of all these contributions should be considered. Indeed, we can show that the metallic resistivity of the NdNiO\textsubscript{3} film over a temperature range of 600 K can be well fit with the DC-PRF of Eqs. \ref{eq for pallaror resistivity}-\ref{definiation of ideal resistivity} with $A$\textsubscript{2} being more than two orders of magnitude smaller than $A$\textsubscript{1} (see Fig. S27 in SI).

One of the interesting features unveiled in Fig. \ref{fig: extracted parameters} is that the increase of $A$\textsubscript{1} saturates at a maximum value $\sim$ 1 $\mu \Omega$ cm/K, which we have previously discussed in relation to the definition of the intermediate region of Fig. \ref{fig:summary of metallic systems}a. However, the DC-PRF allows to extract the linear contribution to resistivity in a wider variety of metallic systems and in a wider temperature range. In this way, we find that the upper limit is, actually, obeyed by all the correlated systems, even in those well-established bad metals.

The same $A$\textsubscript{1} $\sim$1 $\mu \Omega$ cm/K limit has been reported in high-$T$\textsubscript{c} cuprates \cite{hussey2011dichotomy} and has been associated with the PDL. Indeed, we notice that the extracted $A$\textsubscript{1} from those strange metals (inside the yellow-encircled region) approximately approaches this upper limit. Despite being derived for simple and isotropic Fermi surfaces, one can use the Drude formula of conduction to estimate the universal Planckian bound on dissipation ($1/\tau$=$k$\textsubscript{B}T/$\hbar$) and obtain that $A$\textsubscript{1}= ($m^*/n$)($k$\textsubscript{B}/$e^2\hbar$), which includes the carrier density ($n$) and carrier effective mass ($m^*$). This bound is, therefore, system-specific and explains that normal metals, due to their lower $m^*/n$ ratio (see Fig. \ref{fig: extracted parameters}(c) and Supplementary Material Section 3), display smaller $A_1$ values than the correlated metals. Indeed, as shown in the inset of Fig. \ref{fig: extracted parameters}(c), the product $A_1 n/m^*$, which characterizes the PDL, confirms that such a limit is generally obeyed \cite{bruin2013similarity}. Thus, our analysis shows that the Planckian bound is a significant contribution to the $\rho$(\textit{T}) in all investigated metals. The relevance of this limit also in systems that show non-linear-$T$ resistivity \cite{mousatov2021phonons} is then clearly demonstrated using this approach. 

 In contrast, the quadratic $A$\textsubscript{2} coefficient does not display a bound and continues increasing to reach the largest values in bad metals (see Fig. \ref{fig: extracted parameters}(b)). Even though $A$\textsubscript{2} cannot be associated to a unique physical process through all the metal classes, its magnitude reflects the strength of the electron scattering, with the smallest $A$\textsubscript{2} for the normal metals and the largest for the bad metals, as expected.  Interestingly, in most of the investigated bad metals, $\rho$\textsubscript{0} (obtained from the DC-PRF fits) is also significantly larger than in other metals (see Fig. \ref{fig: extracted parameters}(d)), confirming the widespread notion that bad metals are dirty metals \footnote{It should be bare in mind that correlation is not causation, since it is easy to get an ordinary metal with a high $\rho$\textsubscript{0}}. This fact, and the large $A$\textsubscript{2} values, are responsible for the increased $\rho$\textsubscript{sat} values that characterize bad metals. Despite the large uncertainties associated with the values of $\rho$\textsubscript{sat} (\footnote{the parameters resulting from of all the fits and their errors can be found at https://doi.org/10.34894/A1AHZR}), the results of the fits in the systems known as bad metals give rise to ultra-large values of $\rho$\textsubscript{sat}, well above what has been generally considered to be their MIR limit ($<$ 1 m$\Omega$ cm \cite{hussey2004universality,gunnarsson2003colloquium}), as expected in this class of metals.

 However, it is worth to point out, firstly, that the role of the saturation term in the DC-PRF expression is in some cases (i.e. normal metals), that of linearizing the parabolic increase in resistance that arises from Eq. \ref{definiation of ideal resistivity}, in order to reproduce the linear electron-phonon regime at intermediate temperatures; while in some other cases, where the data show signs of saturation, this term can be directly associated to the MIR limit. This further emphasizes the phenomenological character of the  DC-PRF expression. Secondly, the exact magnitude of the MIR limit for electron correlated metals is difficult to determine and it has been proposed that the small Fermi surfaces of these materials can lead to MIR values much larger than the ones earlier proposed \cite{poniatowski2021resistivity}.

Thus, our analysis shows that the different behaviour of $\rho$(\textit{T}) in various systems is mainly determined by the relative importance of $A$\textsubscript{1} and $A$\textsubscript{2}, which can be assessed by the magnitude of $T$\textsuperscript{*}= $A$\textsubscript{1}/$A$\textsubscript{2}, that is the temperature at which the linear and quadratic terms in $\rho$(\textit{T}) become equal \footnote{In a more detailed analysis, this definition could be extended to $T$\textsuperscript{*}/W, where W is the bandwidth of the bands crossing the Fermi energy, so to obtain an adimensional figure of merit to assess the relative relevance of scattering mechanisms},  and their $\rho$\textsubscript{0}. Fig. \ref{fig:  residual resistivity} (a) shows $\rho$\textsubscript{o} as a function of $T$\textsuperscript{*}, evidencing that the bad metals show the smallest values of $T$\textsuperscript{*}; while the strange metals show the largest $T$\textsuperscript{*}. Indeed, the strange metals have a significantly smaller $A$\textsubscript{2}, compared to bad metals, and  similar $A$\textsubscript{1}, which leads to a strongly decreased quadratic contribution, as expected from their close to linear dependence in an extended temperature range. Ayres \textit{et al}. \cite{ayres2021incoherent} proposed that the strange metals host two charge sectors, one containing coherent quasiparticles, and the other one containing scale-invariant ‘Planckian’ dissipators \cite{ayres2021incoherent}. This is well consistent with our findings, with $T$\textsuperscript{*}= $A$\textsubscript{1}/$A$\textsubscript{2} representing the temperature at which one sector takes over the other one.

According to Fig. \ref{fig:  residual resistivity}, there are metals with low $\rho$\textsubscript{0} ($<$ 10\textsuperscript{-3} m$\Omega$ cm), intermediate $\rho$\textsubscript{0} ($\sim$ 10\textsuperscript{-2} m$\Omega$ cm)  and large $\rho$\textsubscript{0} ($\sim$ 1 m$\Omega$ cm) and they can have low $T$\textsuperscript{*} ($<$ 30 K), intermediate $T$\textsuperscript{*}  (30 K$<$ $T$\textsuperscript{*} $<$ 300 K) or large $T$\textsuperscript{*} ($>$ 300 K). Good metals are characterized by a low $\rho$\textsubscript{0}; while bad metals are characterized by a large $\rho$\textsubscript{0}; the rest of the metals (mostly correlated electron systems) display intermediate values of $\rho$\textsubscript{0} and some of them show a large $T$\textsuperscript{*} (see Fig. \ref{fig:  residual resistivity}(a)). These are the strange metals. We have mentioned the difficulties in the classification of some system such as Sr$_2$RuO$_4$, CrO$_2$ or NdNiO$_3$. Our analysis suggest that these materials may be classified as belonging to the intermediate $A$\textsubscript{2} regime of correlated materials and are quite far from the values of bad and good metals (see Fig. \ref{fig:  residual resistivity}(b)). Interestingly, these three materials happen to be unexpectedly clean, according to the $\rho$\textsubscript{0} obtained from the DC-PRF analysis, which are comparable to those of the simple metals, as shown in Fig. \ref{fig: extracted parameters}(d). This observation is an example of the potential benefit of the present proposal to provide a consistent classification of metallic transport on the basis of these two simple parameters.

It is also worth noticing that, in those cases in which $A$\textsubscript{2} $\ll$ $A$\textsubscript{1}, the description of $\rho$\textsubscript{ideal}($T$) with a linear and quadratic term is a very good approximation to the actual data, while larger $A_2$ values, approaching $A_1$, indicate that adding higher order terms in the Taylor expansion is needed in order to better reproduce the data. We have done that in the case of Bi\textsubscript{2}Sr\textsubscript{2}Ca\textsubscript{0.89}Y\textsubscript{0.11}Cu\textsubscript{2}O\textsubscript{y} (Fig. S35 in Supplemental Material), Rb\textsubscript{3}C\textsubscript{60} (Fig. S36 in Supplemental Material), and YBa\textsubscript{2}Cu\textsubscript{3}O\textsubscript{6.45} (Fig. S37 in Supplemental Material) showing that this, indeed, gives better fits. Both the residual resistivity and the linear coefficient are kept unchanged with respect to the original fit. The addition of a higher order term, reduces both $A$\textsubscript{2} and $\rho$\textsubscript{SAT}, emphasizing the phenomenological character of these values, and showing that $A$\textsubscript{2} in Fig. \ref{fig: extracted parameters} encompasses the non-linear contributions to the scattering rate.

\section{CONCLUSION}

The customary classification of metals as normal, bad or strange, runs short to describe the complexity of electron correlated systems, often leading to controversial conclusions. We show that the Hussey formalism applied to cuprates, which consist of $T$-linear ($A_1$) and $T$-quadratic ($A_2$) components added to the residual resistivity and in parallel with a saturation term, can describe the behaviour of metals far more generally, offering the opportunity for a unified description. Moreover, we also showed that defining a temperature $T$\textsuperscript{*}= $A_1$/$A_2$ may provide a general framework to classify metals in accordance to the relative magnitude of $\rho$\textsubscript{0} and $T$\textsuperscript{*}. In strange metals, $T$\textsuperscript{*} describes the crossover from coherent quasi-particle scattering to scale-invariant "Planckian" dissipation. Generally, $A_1$ and $A_2$ do not have a well-defined physical origin and they simply represent the incoherent dissipation and the coherent quasi-particle scattering contributions to the resistivity, respectively. $A_1$ is found to reach an upper bound, for sufficiently large ($m^*/n$) ratios. The clear link of this bound with the Planckian dissipation limit supports its proposed universality \cite{bruin2013similarity,legros2019universal}, extending its scope to a larger number of metals and evidencing that all metals may obey the Planckian constrain.
  

\section{Acknowledgements}
We are indebted to Nigel Hussey, Jan Zaanen, Thom Palstra and Francisco Rivadulla for insightful discussions. We are grateful to Jacob Baas, Arjun Joshua and Henk Bonder for their invaluable technical support. Qikai Guo acknowledges financial support from a China Scholarship Council (CSC) grant and we both acknowledge financial support from the Ubbo Emmius Funds (University of Groningen).

\section{Competing interests}

The Authors declare no Competing Financial or Non-Financial Interests.

\section{Data Availability}
The data that support the findings of this study are available from the corresponding authors upon reasonable request.

\section{Author Contributions}
Q.G. and B.N. designed the project; Q.K. synthesized the nickelate films and performed the transport measurements. C.M. performed the electron microscopy measurements and analysed the data. Q.G. collected all the reported data and performed the fits and analysis with feedback from all the authors. Q.G. and B.N. wrote the first version of the manuscript. All the authors discussed the results, read and contributed to the different versions of the manuscript.

\clearpage

\section{Supplementary Material}

\subsection{Characterization of NdNiO\textsubscript{3} films}

The crystalline structure of the NdNiO\textsubscript{3} film was studied by atomic resolution scanning transmission electron microscopy (STEM). A high-angle annular dark field (HAADF) STEM image shown in Fig. \ref{fig:STEM}a indicates a high crystalline quality in the NNO films with atomically sharp interface with the substrate. No misfit dislocations or other common defects such as Ruddlesden-Popper (RP) faults, which are known to form in nickelates in the presence of excess $A$-cations or oxygen vacancies, are observed in the lattice.  

The annular bright field (ABF) image displayed in Fig. \ref{fig:STEM}b provides further evidence of the octahedral oxygen rotations of the NNO orthorhombic phase by direct visualization of the oxygen lattice along the (110) orientation.

\renewcommand{\thefigure}{S1}
\begin{figure}[htp!]
\centering
\includegraphics[width=0.5\textwidth]{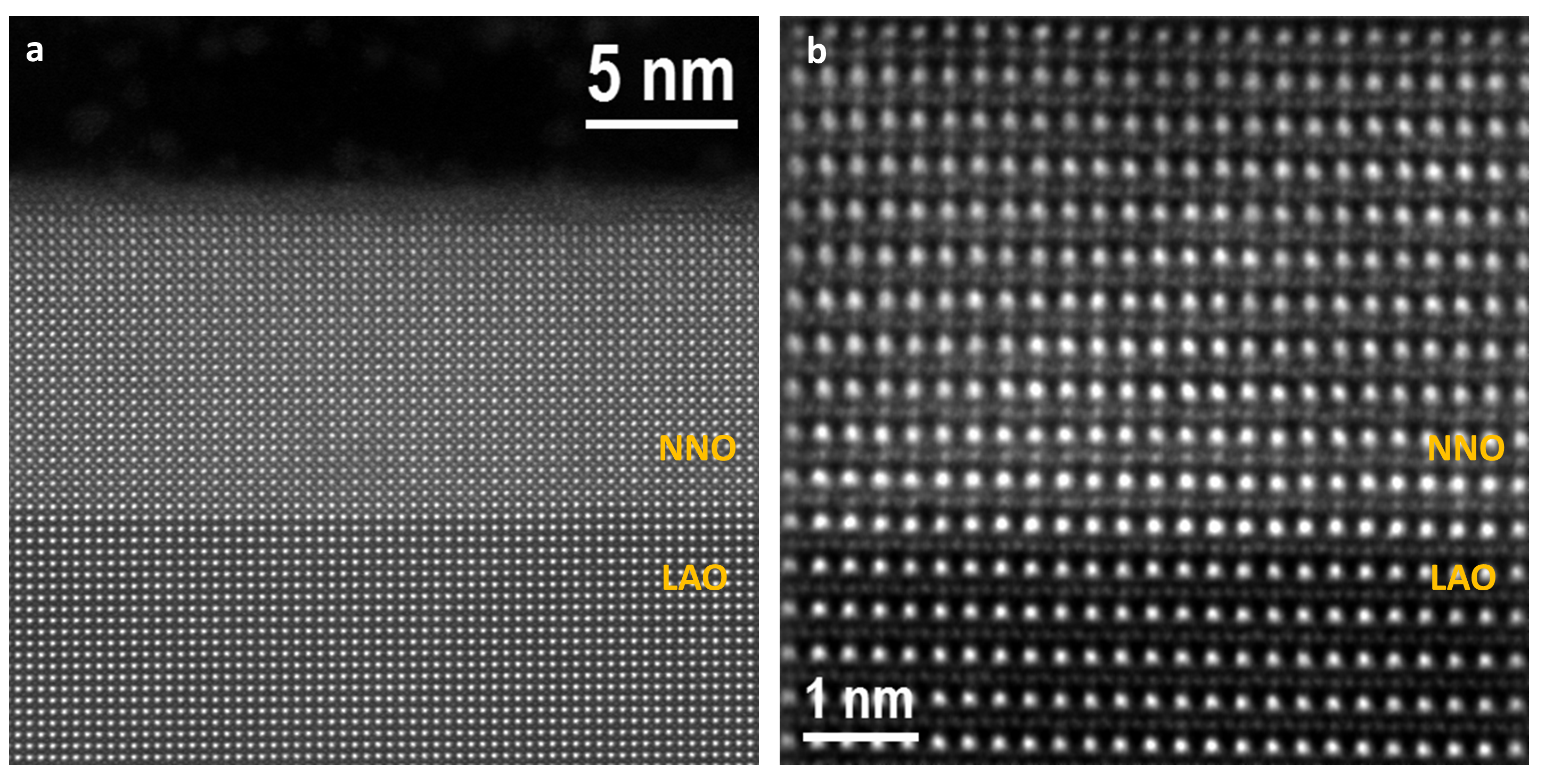}
\caption{\textbf{Characterization of NdNiO\textsubscript{3} film.} \textbf{a} Cross-sectional HAADF-STEM images of a 10-nm-thick NNO film grown on a LAO (001) substrate observed along a (100) crystal orientation. \textbf{b} Cross-sectional ABF-STEM image  with inverted contrast of the same film observed along the (110) crystal orientation where octahedra oxygen rotations of the orthorhombic NNO crystal structure can be observed.}
\label{fig:STEM}
\end{figure}

\subsection{Fit to various metals}

In the present work, a parallel resistor formalism \cite{cooper2009anomalous}:  1/$\rho(T)$=1/($\rho_{0}$+AT+$BT^2$)+1/$\rho \textsubscript{sat}$, was employed to describe the electrical resistivity of NdNiO\textsubscript{3} film. For comparison, the resistivity data of various metallic systems from simple metals to strongly correlated metals were extracted from previous works. In the following, we will discuss in detail the fits to all the materials using the same approach.

\subsection{La\textsubscript{2-x}Sr\textsubscript{x}CuO\textsubscript{4}}

A systematical analysis had been performed on the electrical resistivity of overdoped (x$>$0.15) La\textsubscript{2-x}Sr\textsubscript{x}CuO\textsubscript{4} by Cooper et al. \cite{cooper2009anomalous}. Herein, we extracted the resistivity data from ref. \cite{cooper2009anomalous} and performed the fit in the same way as described in the main text for the NdNiO\textsubscript{3} film. It is worth to emphasize that the $\rho$\textsubscript{sat} is a free parameter in our fits, while a fixed value (900 $\mu \Omega$ cm) of $\rho$\textsubscript{sat} is employed in Cooper's work. As shown in the figures below, the resistivity curves of all the reported doping levels by Cooper et al. are well reproduced by our analysis. The parameters used in the fit are shown in the inset of each figure. Interestingly, we found that the extracted $\rho$\textsubscript{sat} from the fits show a decrease with increasing hole doping levels. For those with overdoping (x $>$ 0.18), the values of $\rho$\textsubscript{sat} are almost comparable with the fixed value used by Cooper et al. However, the fits corresponding to the samples with nearly optimized doping (p=0.17 and 0.18), which show nearly linear-$T$ resistivity, give rise to a significantly larger $\rho$\textsubscript{sat}. In addition we have also used resistivity of LSCO with x=0.15 extracted from the work of Takagi et al \cite{takagi1992systematic}, whose sample show an even longer linear regime. The data were analyzed with the same criterion, as shown in Fig. \ref{fig:LSCO15}. In this case, the linear-$T$-resistivity persists up to almost 1000 K and can even continue further, as suggested by the large value of predicted $\rho$\textsubscript{sat}. 
. 

The extracted $A$\textsubscript{1} and $A$\textsubscript{2} as a function of hole doping levels is shown in Figure \ref{fig:Extracted coefficients}. Notably, both marked kinks (solid dashed line) in $A$\textsubscript{1} and $A$\textsubscript{2} (corresponding to the $\alpha$\textsubscript{1} and $\alpha$\textsubscript{2} in their work, respectively) and also the evolution of these two coefficients with hole doping, reported by Cooper et al., are well reproduced by our fit. We believe that all these features demonstrate a reliability of our approach.

\renewcommand{\thefigure}{S2}
\begin{figure}[htp!]
\centering
\includegraphics[width=0.4\textwidth]{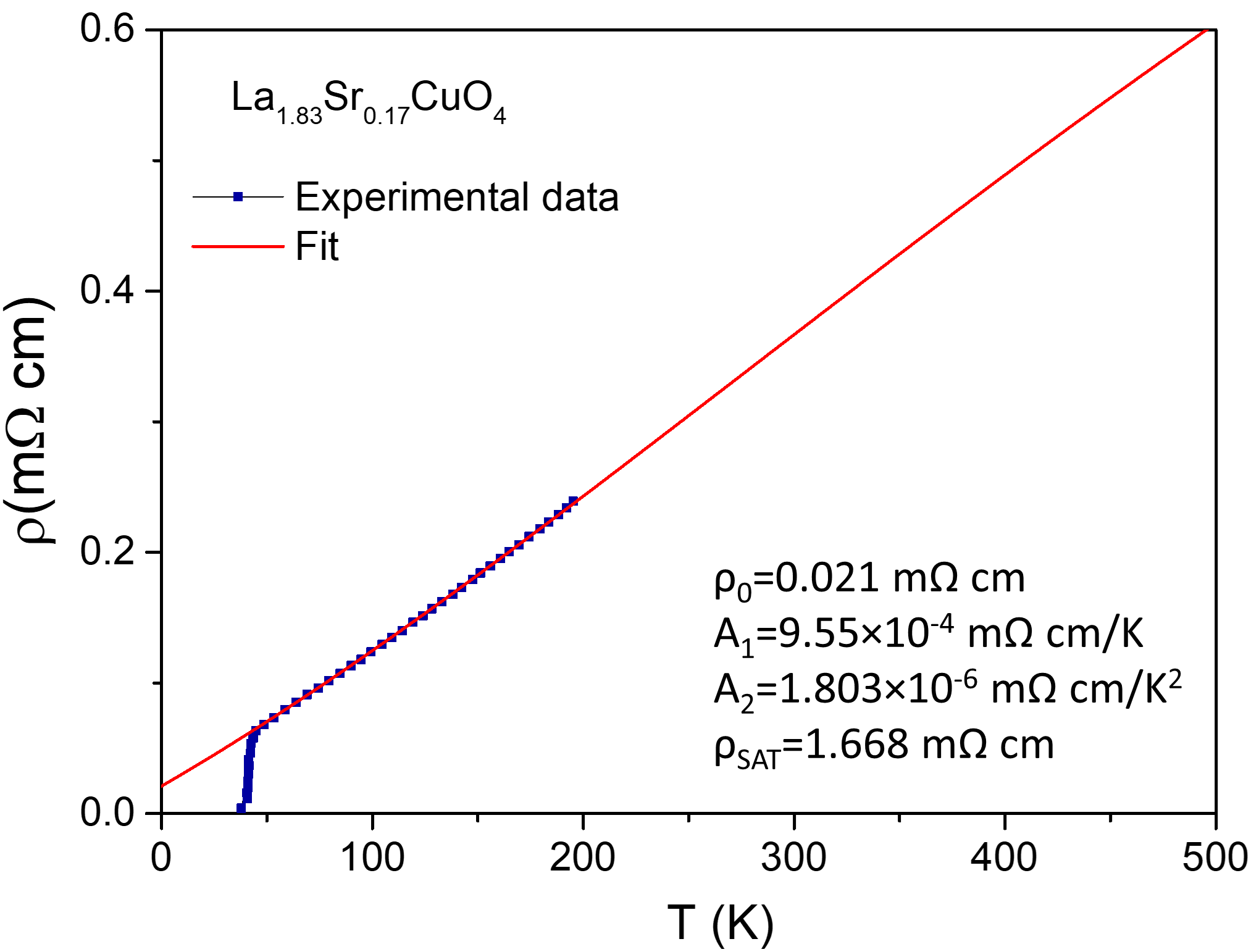}
\caption{Fitting of $\rho$(T) of La\textsubscript{1.83}Sr\textsubscript{0.17}CuO\textsubscript{4}. Data from Ref. \cite{cooper2009anomalous}.}
\label{fig:LSCO17}
\end{figure}

\renewcommand{\thefigure}{S3}
\begin{figure}[htp!]
\centering
\includegraphics[width=0.4\textwidth]{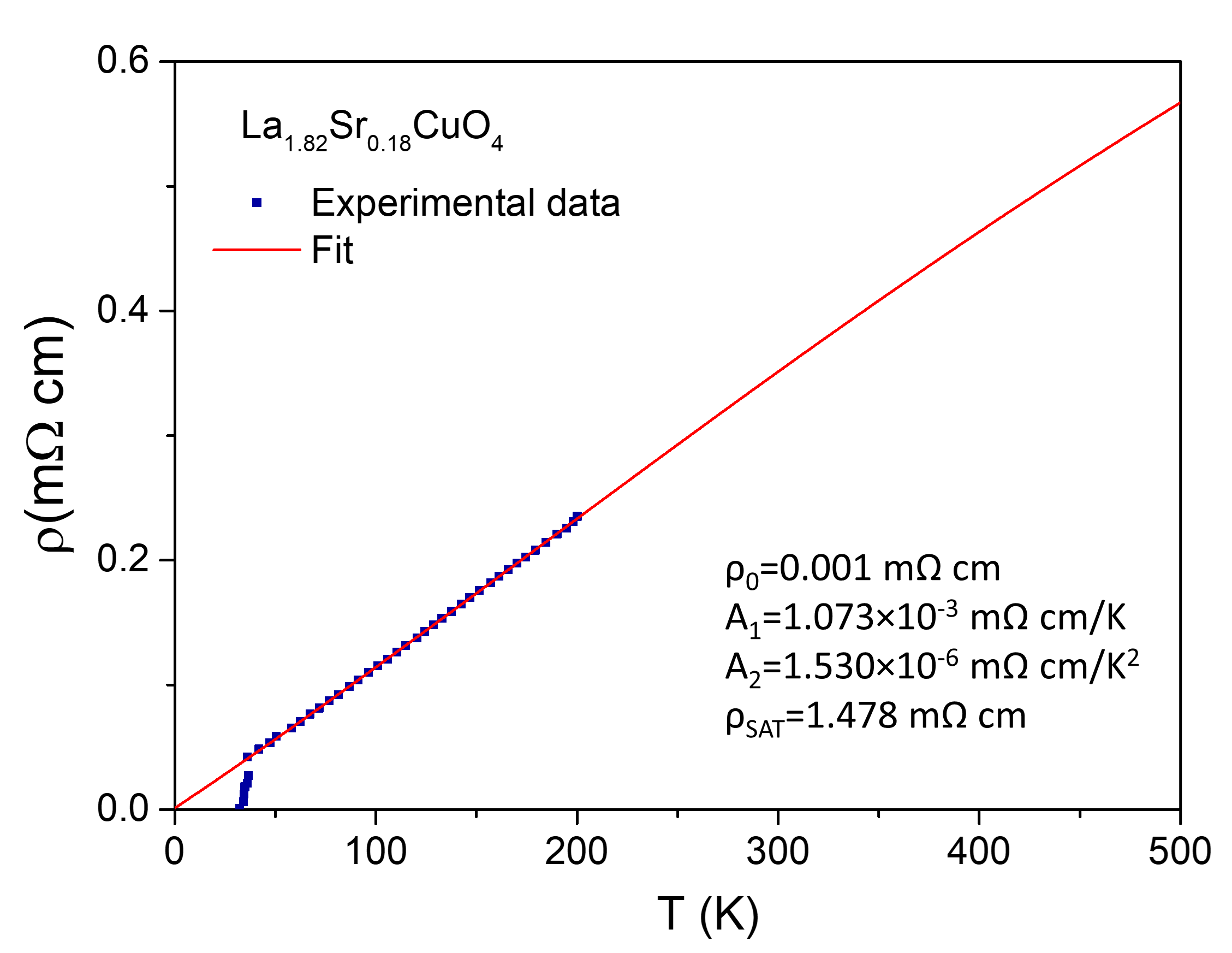}
\caption{Fitting of $\rho$(T) of La\textsubscript{1.82}Sr\textsubscript{0.18}CuO\textsubscript{4}. Data from Ref. \cite{cooper2009anomalous}.}
\label{fig:LSCO18}
\end{figure}

\renewcommand{\thefigure}{S4}
\begin{figure}[htp!]
\centering
\includegraphics[width=0.4\textwidth]{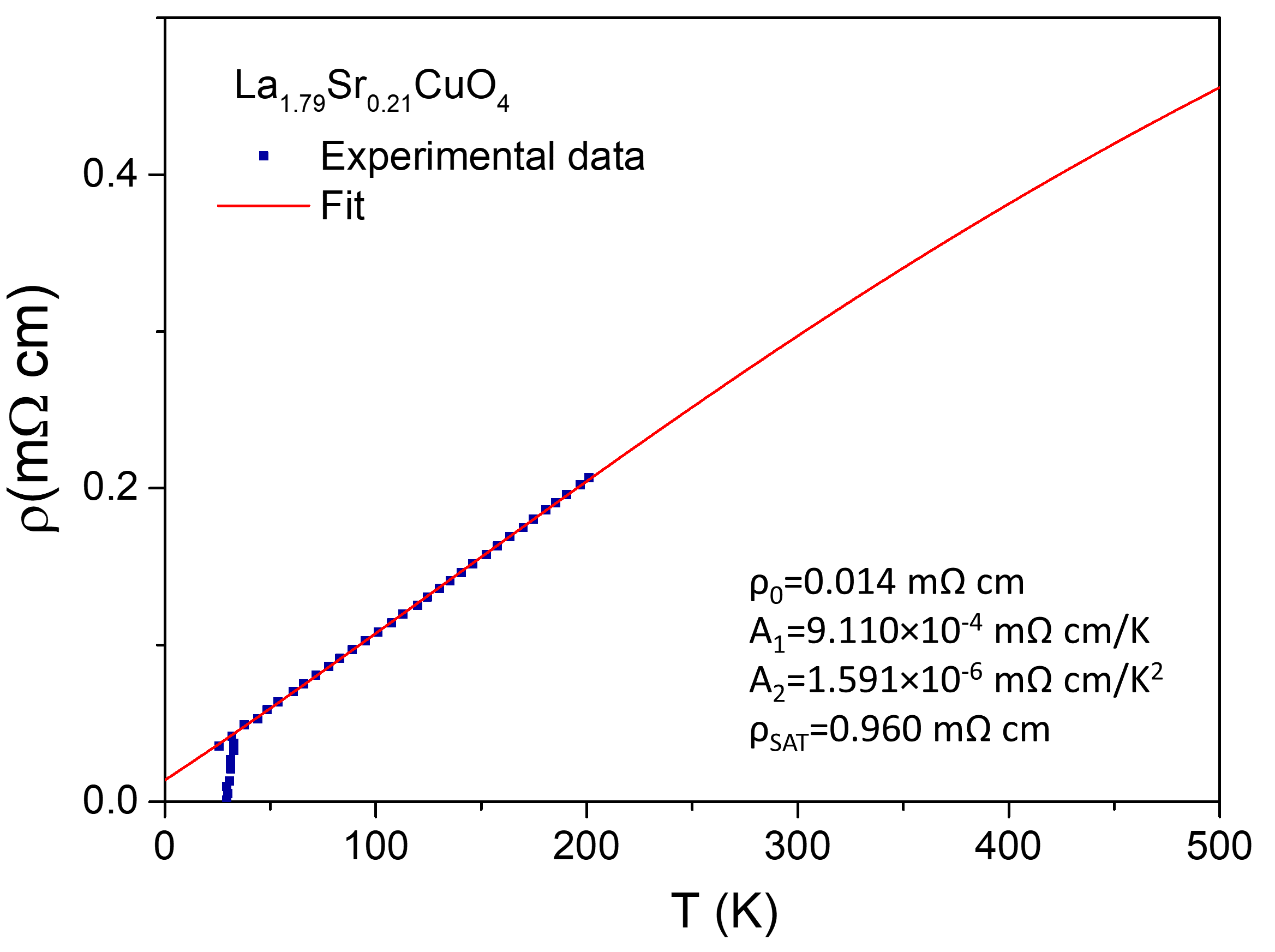}
\caption{Fitting of $\rho$(T) of La\textsubscript{1.79}Sr\textsubscript{0.21}CuO\textsubscript{4}. Data from Ref. \cite{cooper2009anomalous}.}
\label{fig:LSCO21}
\end{figure}

\renewcommand{\thefigure}{S5}
\begin{figure}[htp!]
\centering
\includegraphics[width=0.4\textwidth]{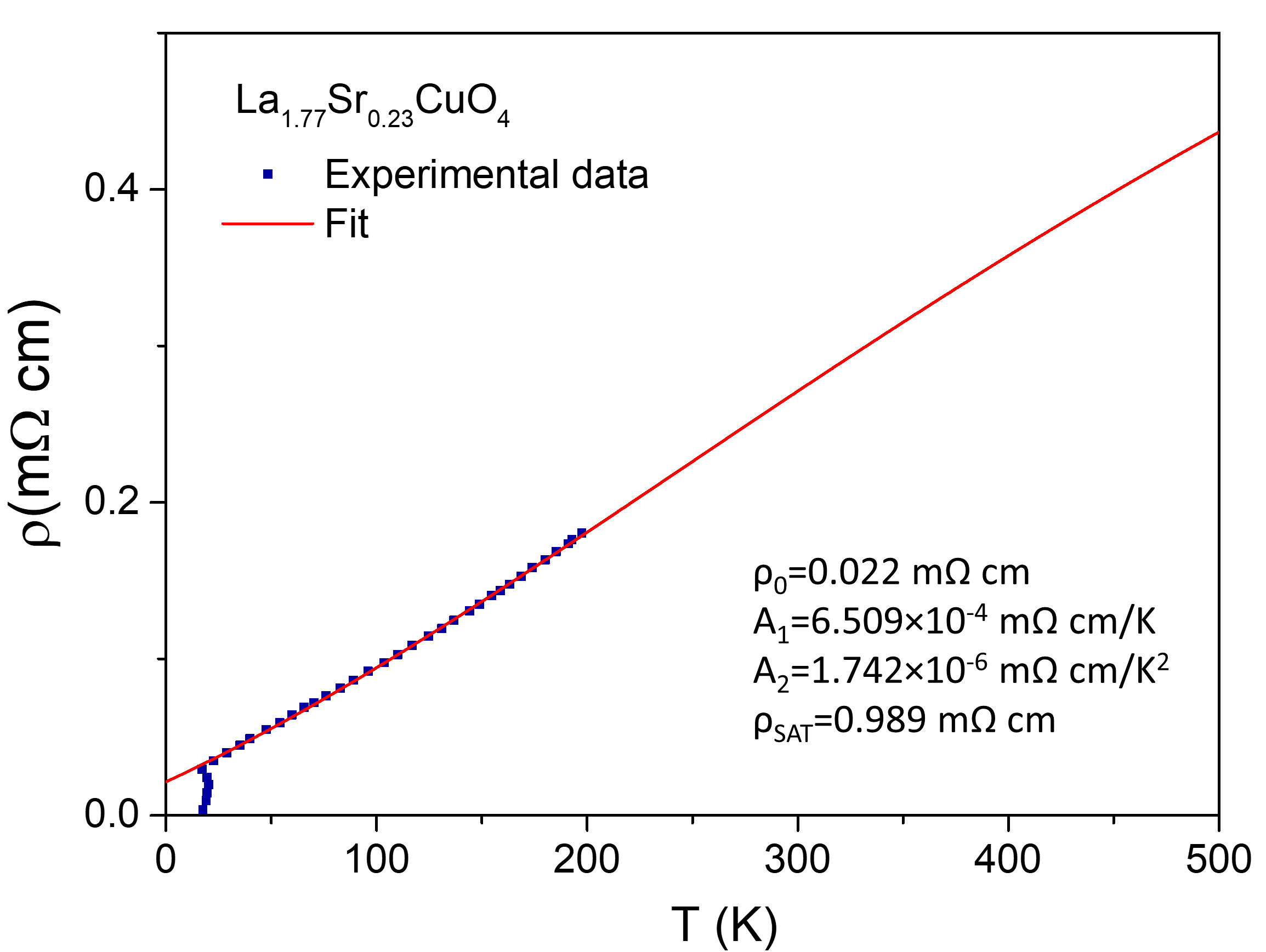}
\caption{Fitting of $\rho$(T) of La\textsubscript{1.77}Sr\textsubscript{0.23}CuO\textsubscript{4}. Data from Ref. \cite{cooper2009anomalous}.}
\label{fig:LSCO23}
\end{figure}

\renewcommand{\thefigure}{S6}
\begin{figure}[htp!]
\centering
\includegraphics[width=0.4\textwidth]{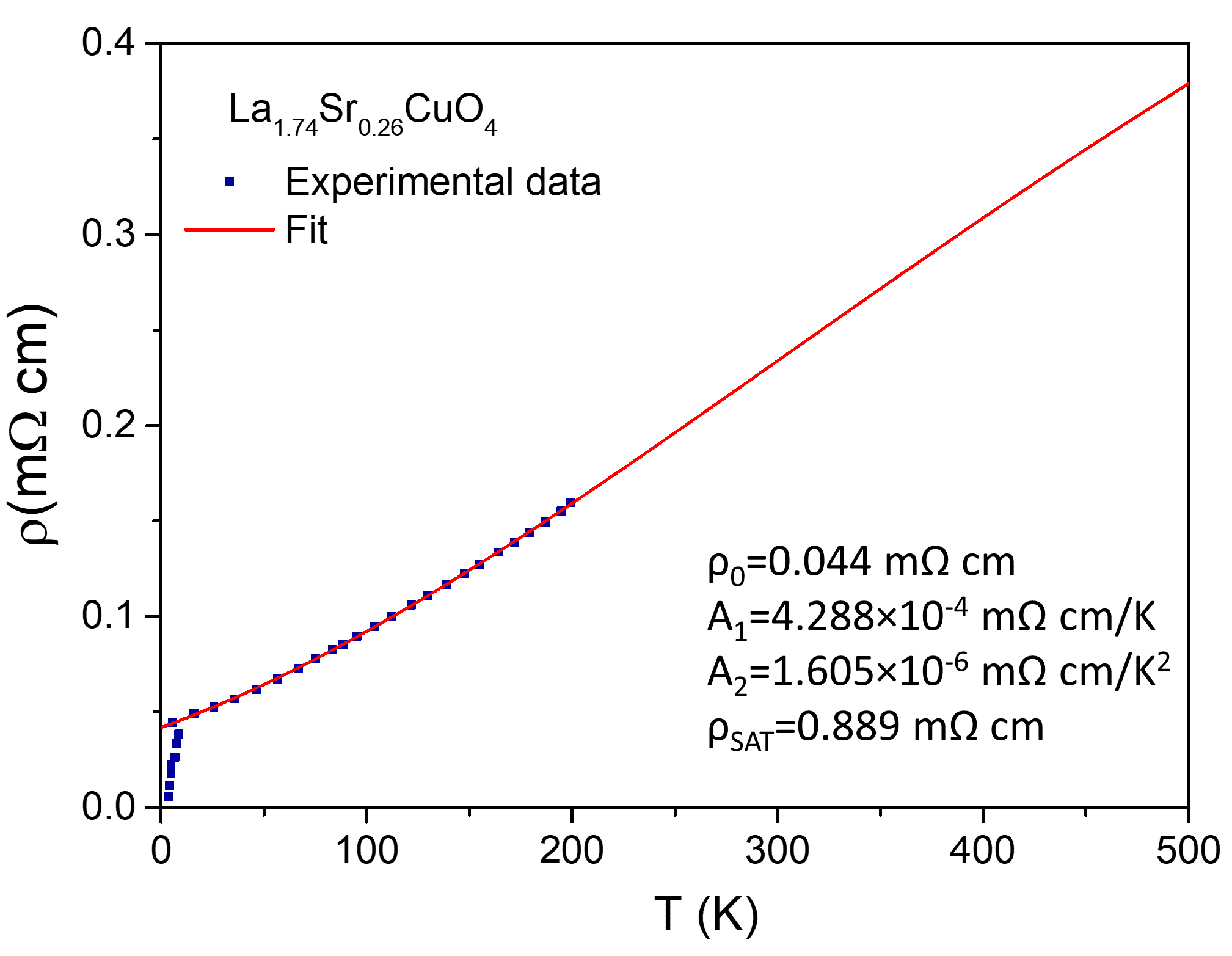}
\caption{Fitting of $\rho$(T) of La\textsubscript{1.74}Sr\textsubscript{0.26}CuO\textsubscript{4}. Data from Ref. \cite{cooper2009anomalous}.}
\label{fig:LSCO26}
\end{figure}

\renewcommand{\thefigure}{S7}
\begin{figure}[htp!]
\centering
\includegraphics[width=0.4\textwidth]{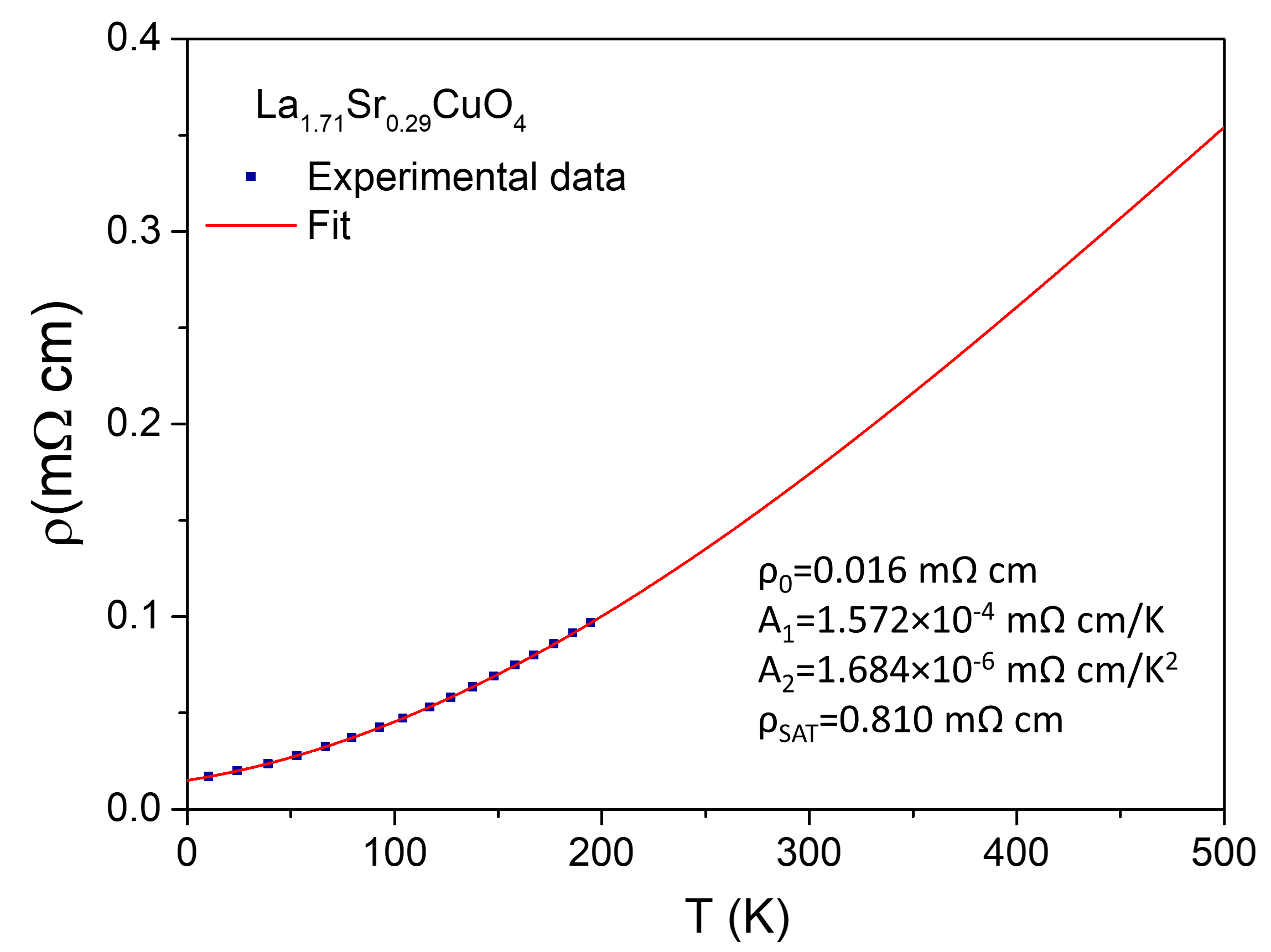}
\caption{Fitting of $\rho$(T) of La\textsubscript{1.71}Sr\textsubscript{0.29}CuO\textsubscript{4}. Data from Ref. \cite{cooper2009anomalous}.}
\label{fig:LSCO29}
\end{figure}

\renewcommand{\thefigure}{S8}
\begin{figure}[htp!]
\centering
\includegraphics[width=0.4\textwidth]{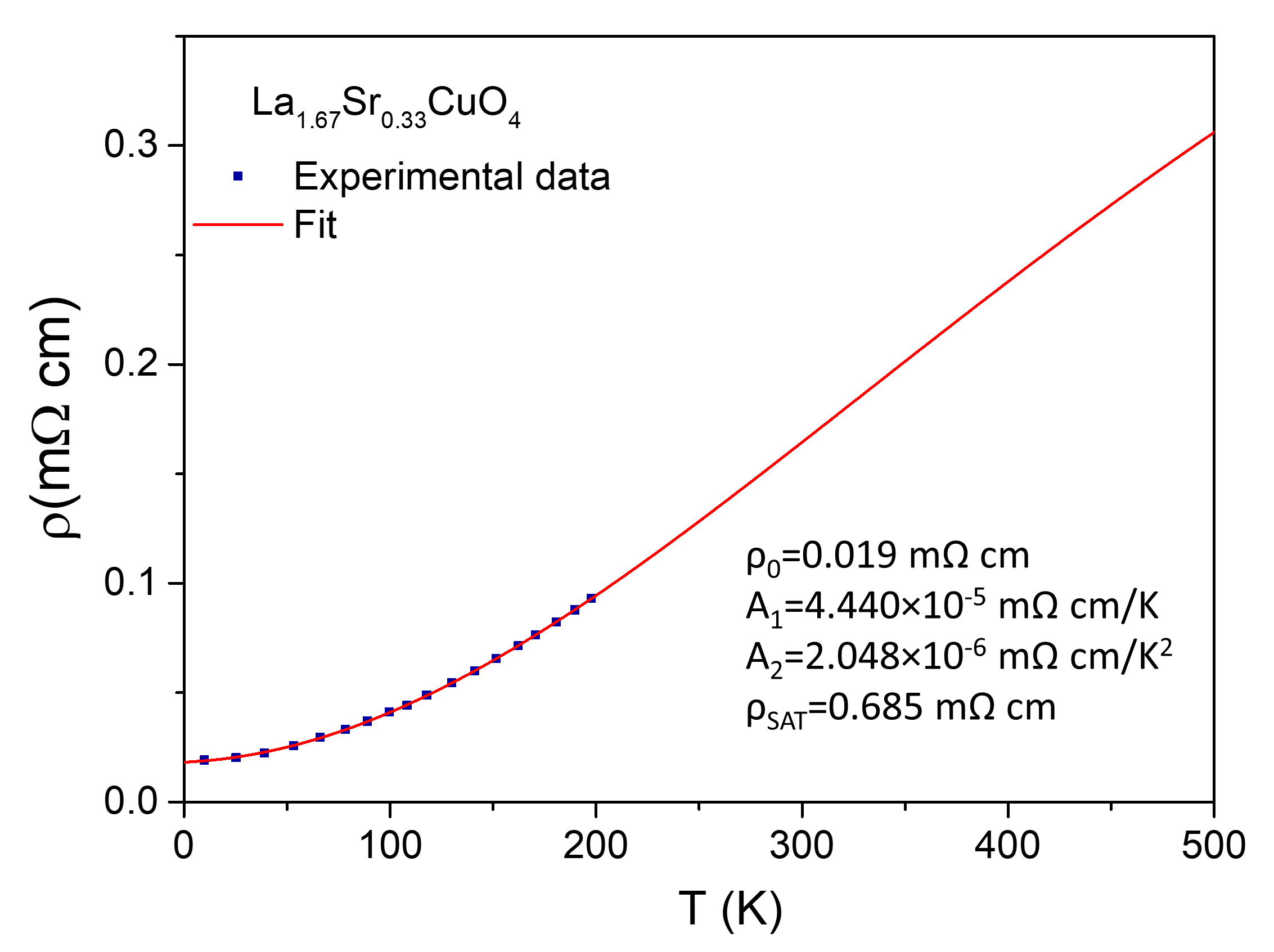}
\caption{Fitting of $\rho$(T) of La\textsubscript{1.67}Sr\textsubscript{0.33}CuO\textsubscript{4}.Data from Ref. \cite{cooper2009anomalous}.}
\label{fig:LSCO33}
\end{figure}

\renewcommand{\thefigure}{S9}
\begin{figure}[htp!]
\centering
\includegraphics[width=0.4\textwidth]{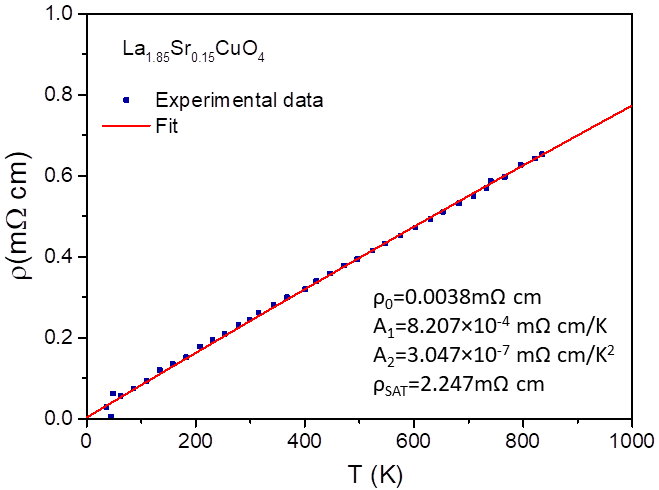}
\caption{Fitting of $\rho$(T) of La\textsubscript{1.85}Sr\textsubscript{0.15}CuO\textsubscript{4}. Data were extracted from Ref. \cite{takagi1992systematic}}
\label{fig:LSCO15}
\end{figure} 

\renewcommand{\thefigure}{S10}
\begin{figure}[htp!]
\centering
\includegraphics[width=0.4\textwidth]{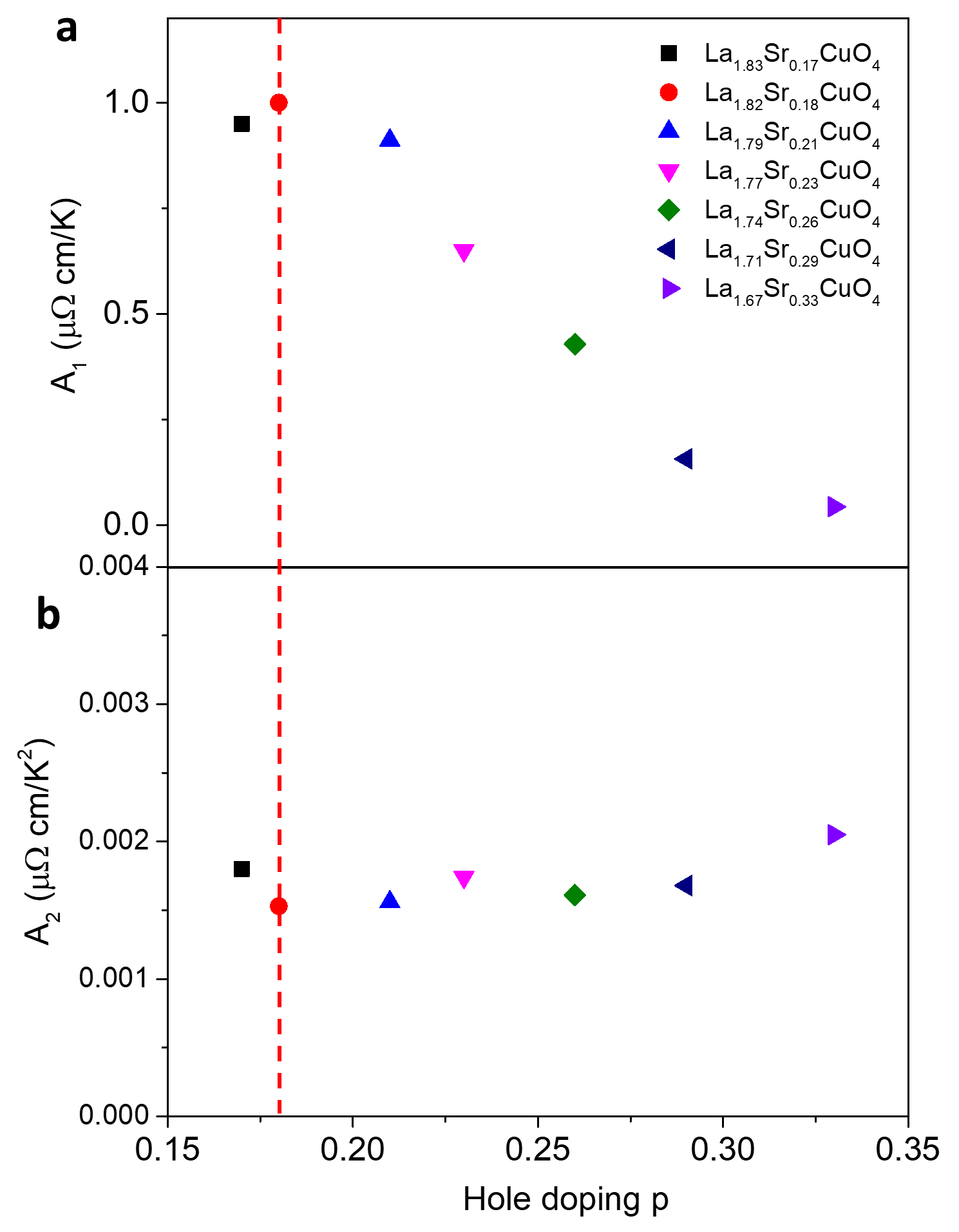}
\caption{Extracted coefficients \textbf{a} $A$\textsubscript{1} and \textbf{b} $A$\textsubscript{2} from the fit to the resistivity of La\textsubscript{2-x}Sr\textsubscript{x}CuO\textsubscript{4} in Ref. \cite{cooper2009anomalous} as a function of hole doping.}
\label{fig:Extracted coefficients}
\end{figure}

\subsection{Bad metals}

Bad metals are characterized by their large resistivity, which can increase across the predicted Mott-Ioffe-Regel (MIR) limit even at low temperatures. In the present work, resistivity data of four different systems, such as underdoped La\textsubscript{2-x}Sr\textsubscript{x}CuO\textsubscript{4}, Bi\textsubscript{2}Sr\textsubscript{2}Ca\textsubscript{0.89}Y\textsubscript{0.11}Cu\textsubscript{2}O\textsubscript{y}, Rb\textsubscript{3}C\textsubscript{60}, YBa\textsubscript{2}Cu\textsubscript{3}O\textsubscript{6.45}, were extracted from the refs. \cite{takagi1992systematic,wang1996observation,hebard1993absence,ito1993systematic}, respectively. Among them, the resistivity of La\textsubscript{1.96}Sr\textsubscript{0.04}CuO\textsubscript{4}, La\textsubscript{1.93}Sr\textsubscript{0.07}CuO\textsubscript{4}, and Bi\textsubscript{2}Sr\textsubscript{2}Ca\textsubscript{0.89}Y\textsubscript{0.11}Cu\textsubscript{2}O\textsubscript{y} show obvious saturation with values well above the MIR limit. However, both Rb\textsubscript{3}C\textsubscript{60} and YBa\textsubscript{2}Cu\textsubscript{3}O\textsubscript{6.45} display a continuously increased resistivity in the whole investigated temperature range with no sign of saturation. Despite of this different performance, the fit with the parallel resistor model shows to describe the experimental data adequately over a wide temperature range.

\renewcommand{\thefigure}{S11}
\begin{figure}[htp!]
\centering
\includegraphics[width=0.4\textwidth]{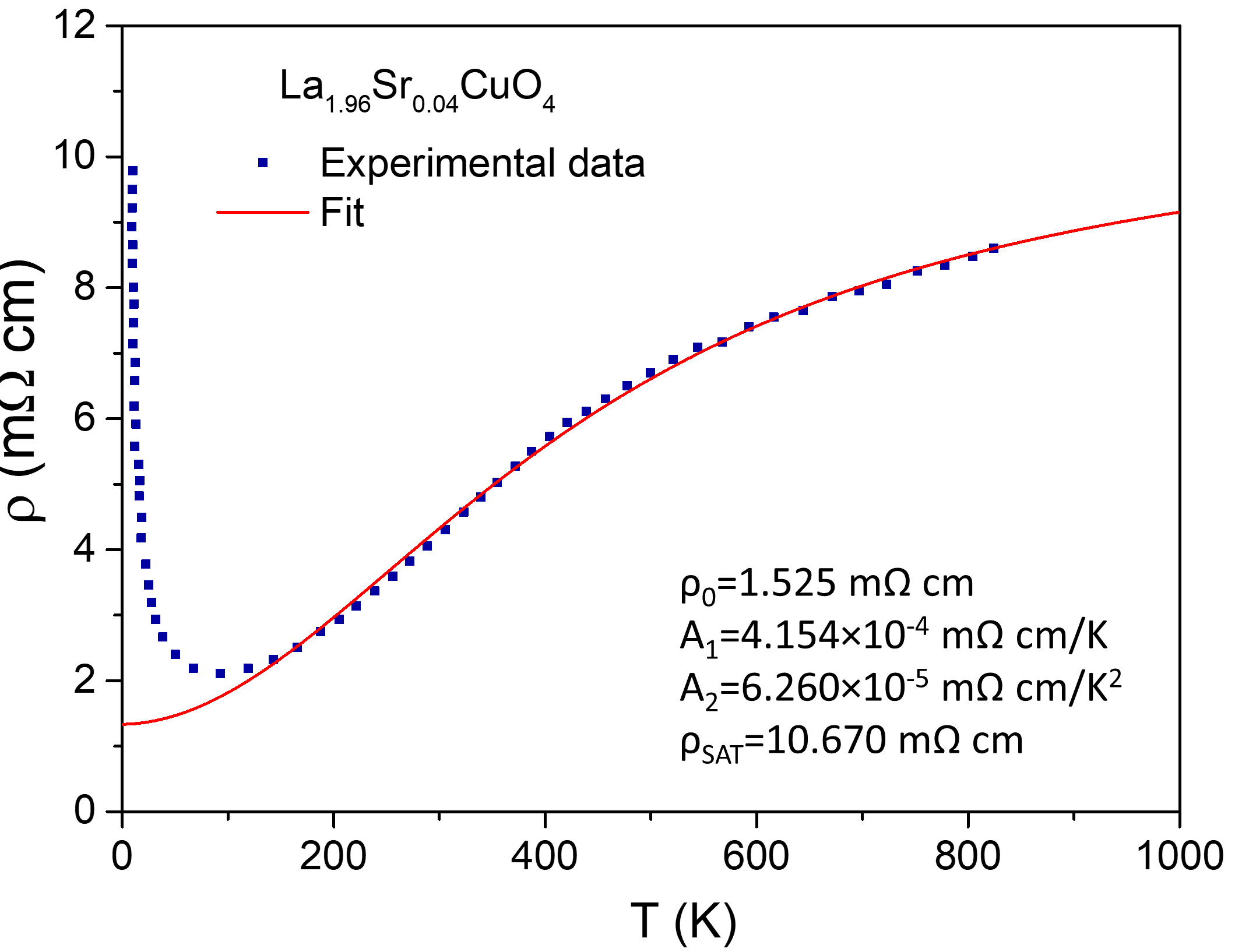}
\caption{Fitting of $\rho$(T) of La\textsubscript{1.96}Sr\textsubscript{0.04}CuO\textsubscript{4}. Data from Ref.\cite{takagi1992systematic}.}
\label{fig:metallic resistivity}
\end{figure} 

\renewcommand{\thefigure}{S12}
\begin{figure}[htp!]
\centering
\includegraphics[width=0.4\textwidth]{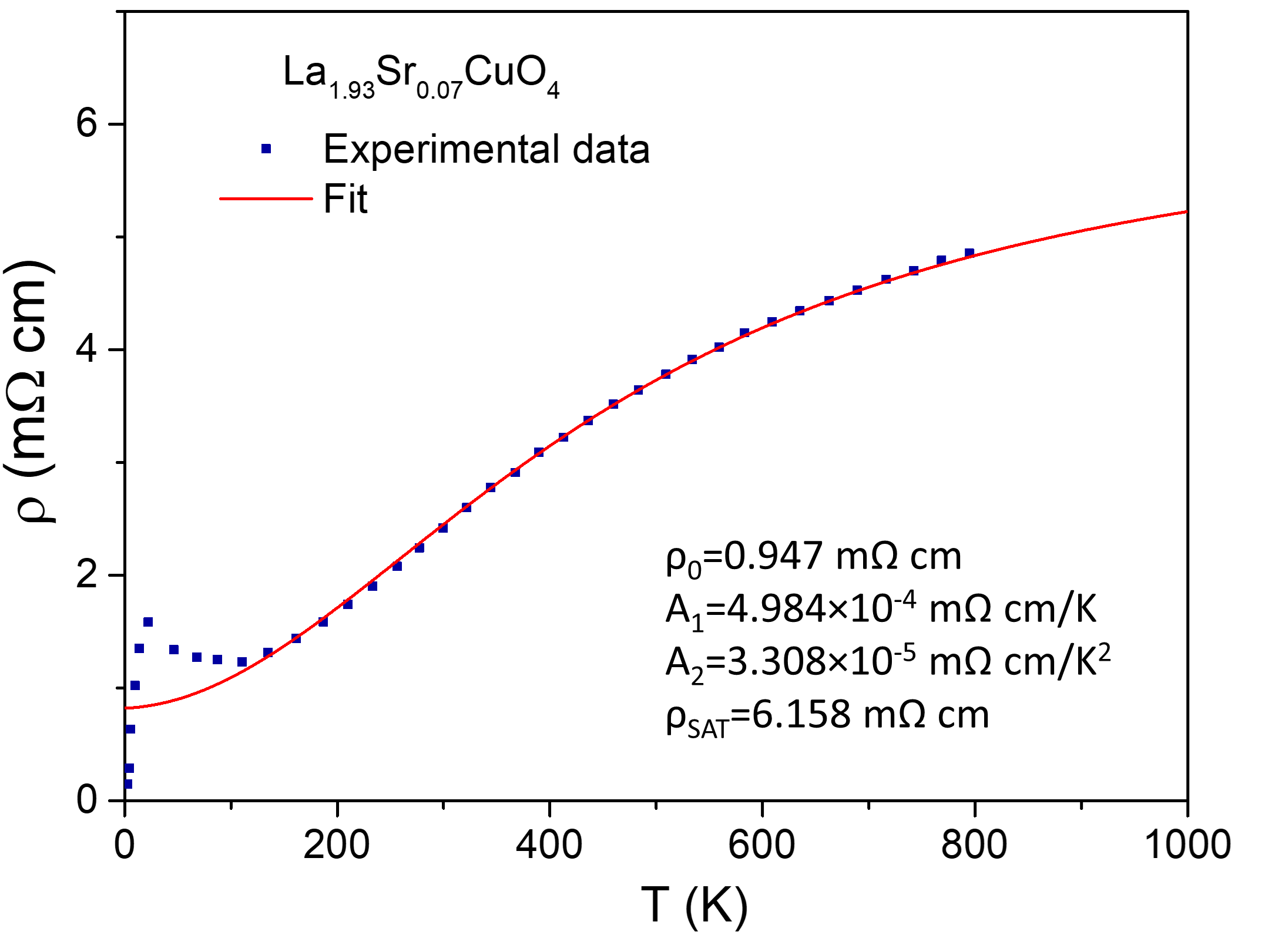}
\caption{Fitting of $\rho$(T) of La\textsubscript{1.93}Sr\textsubscript{0.07}CuO\textsubscript{4}. Data from Ref.\cite{takagi1992systematic}.}
\label{fig:metallic resistivity}
\end{figure}

\renewcommand{\thefigure}{S13}
\begin{figure}[htp!]
\centering
\includegraphics[width=0.4\textwidth]{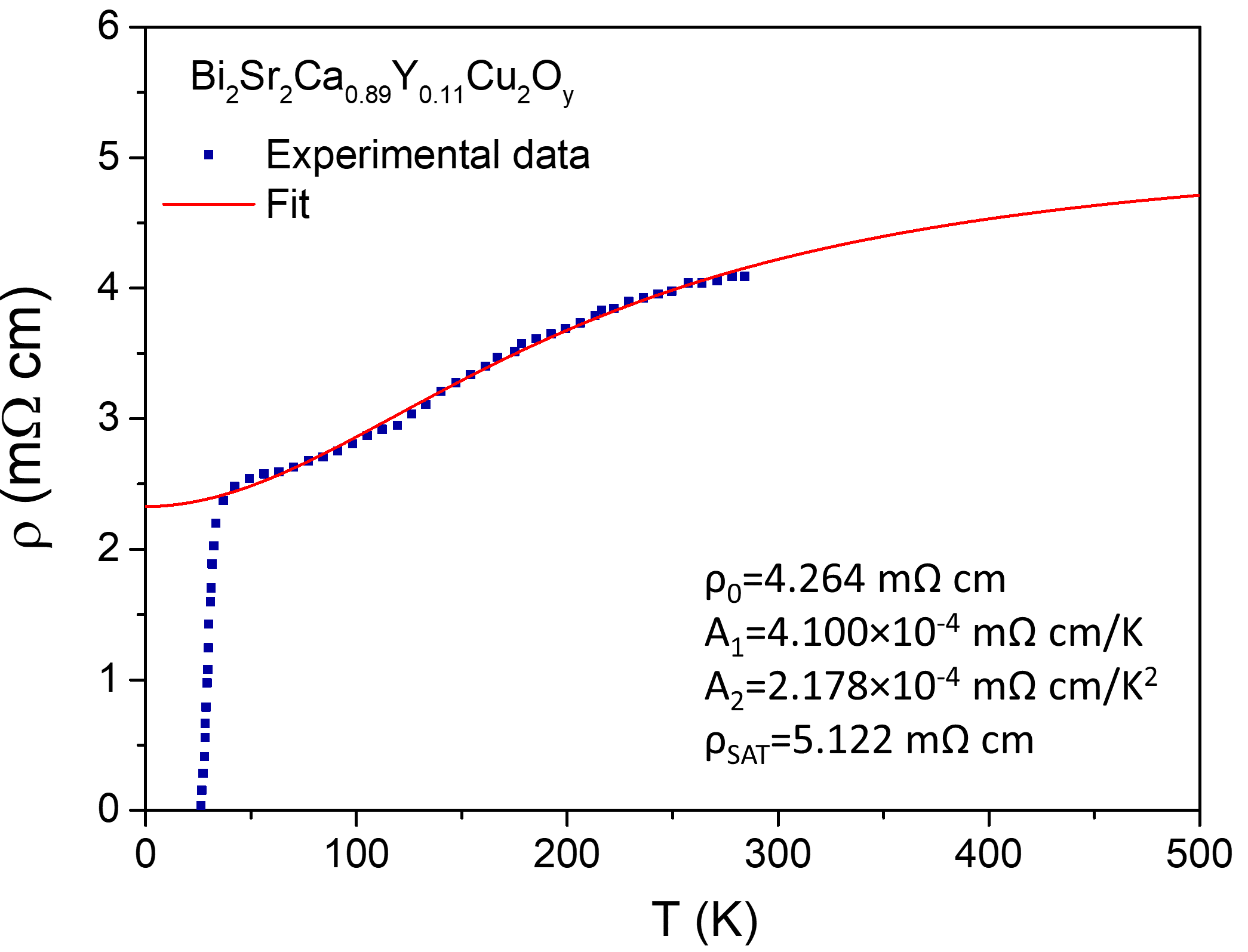}
\caption{Fitting of $\rho$(T) of Bi\textsubscript{2}Sr\textsubscript{2}Ca\textsubscript{0.89}Y\textsubscript{0.11}Cu\textsubscript{2}O\textsubscript{y}. Data from Ref.\cite{wang1996observation}. }
\label{fig:metallic resistivity}
\end{figure}

\renewcommand{\thefigure}{S14}
\begin{figure}[htp!]
\centering
\includegraphics[width=0.4\textwidth]{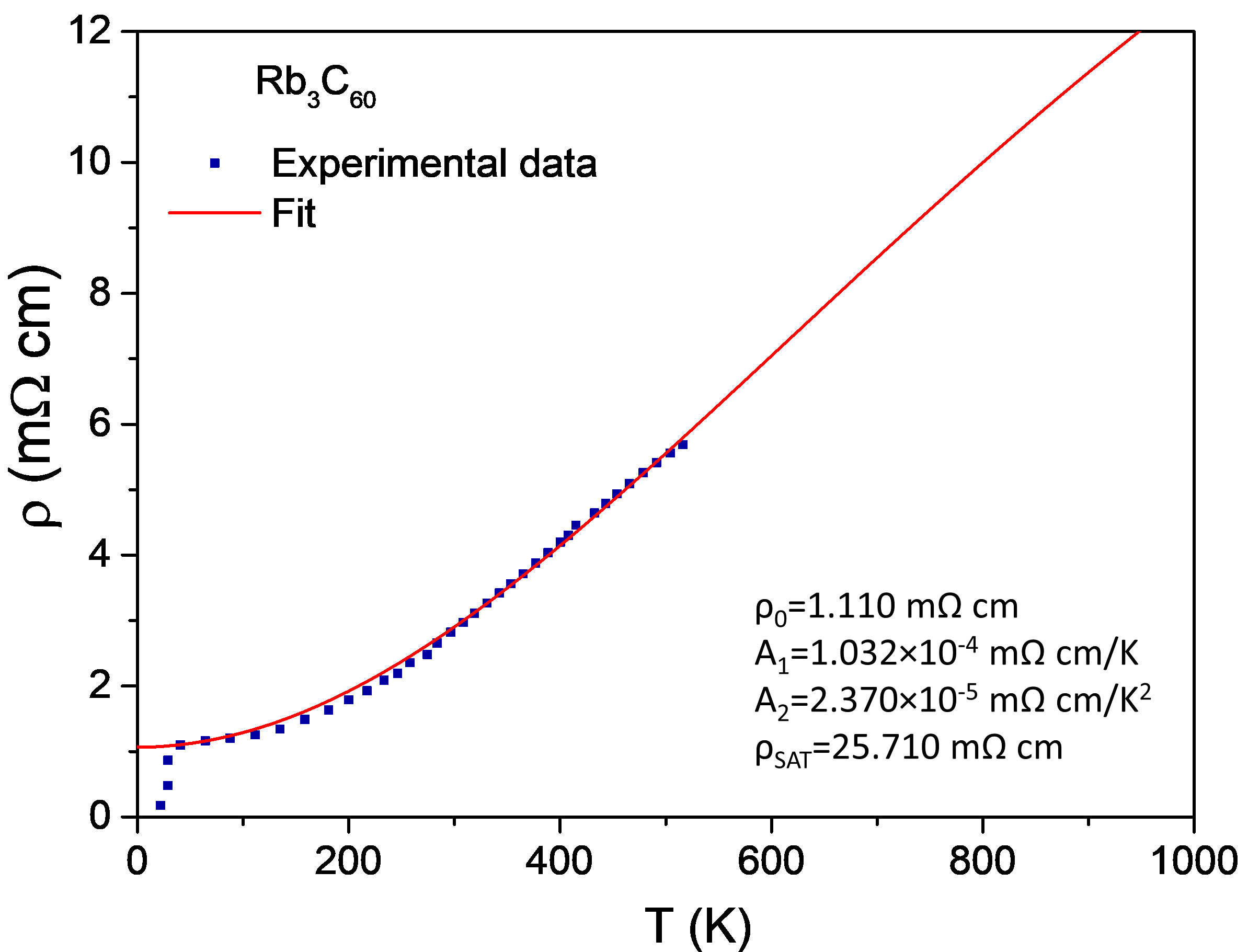}
\caption{Fitting of $\rho$(T) of Rb\textsubscript{3}C\textsubscript{60}. Data from Ref.\cite{hebard1993absence}.}
\label{fig:metallic resistivity}
\end{figure}

\renewcommand{\thefigure}{S15}
\begin{figure}[htp!]
\centering
\includegraphics[width=0.4\textwidth]{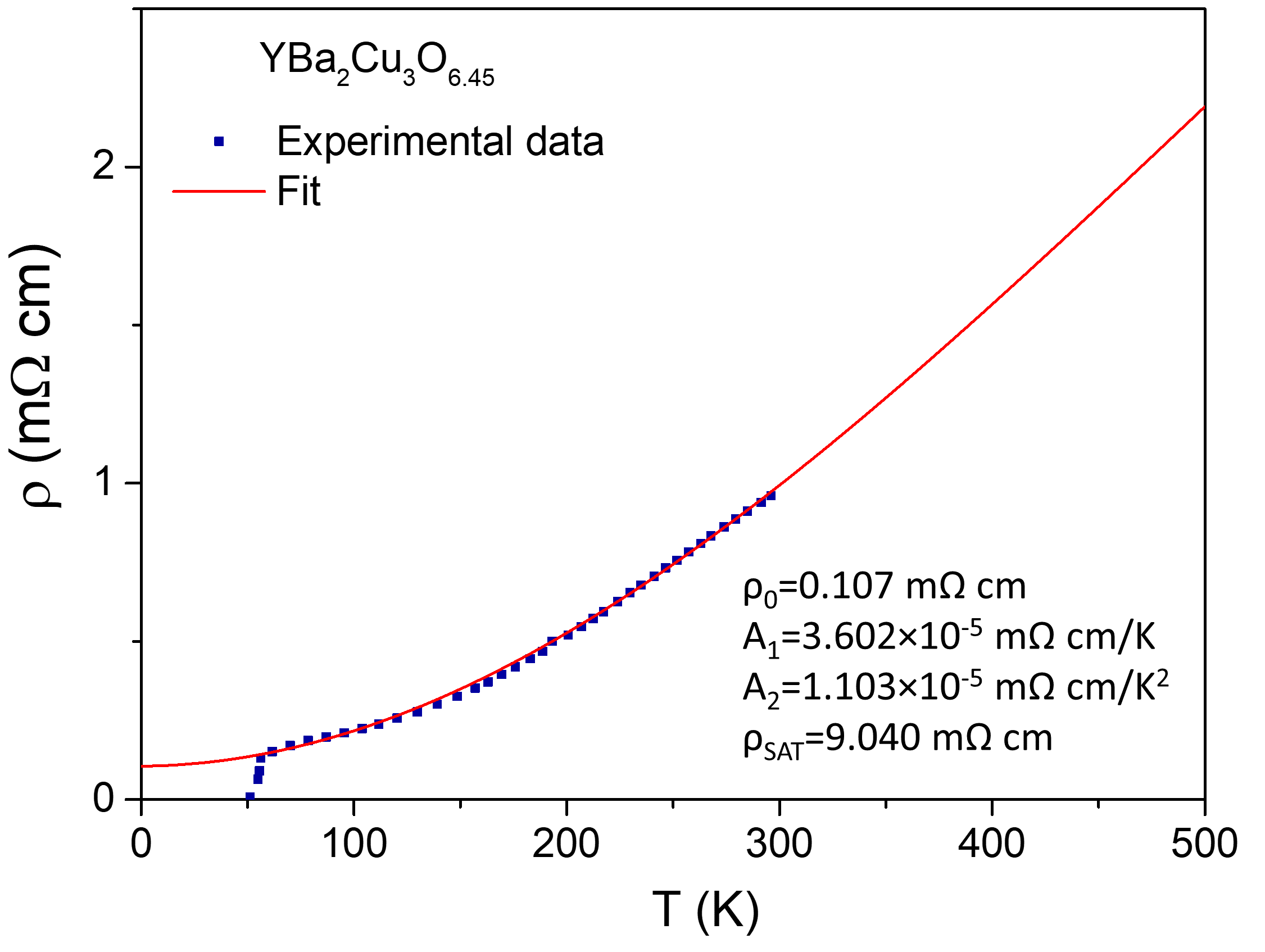}
\caption{Fitting of $\rho$(T) of YBa\textsubscript{2}Cu\textsubscript{3}O\textsubscript{6.45}. Data from Ref.\cite{ito1993systematic}.}
\label{fig:metallic resistivity}
\end{figure}

\subsection{Other correlated metals}

\begin{itemize}

\item \textbf{Sr\textsubscript{2}RuO\textsubscript{4}}

An extended measurement of resistivity in Sr\textsubscript{2}RuO\textsubscript{4} had been performed by Tyler et al. \cite{tyler1998high}. This material displays an interesting case among strongly correlated metals. Sr\textsubscript{2}RuO\textsubscript{4} has been proven to be a very good metal at low temperature, following to a Fermi-liquid quasi-particle scenario. However, the increase of resistivity at high-$T$ shows no sign of saturation at the Mott-Ioffe-Regel limit, invaliding the quasi-particle description. Indeed, the fit of $\rho$(T) extracted from ref. \cite{tyler1998high} gave a saturation resistivity above 7.4 $\mu \Omega$ cm, which is far beyond its calculated MIR limit ($\sim$ 0.2 $\mu \Omega$ cm).

\renewcommand{\thefigure}{S16}
\begin{figure}[htp!]
\centering
\includegraphics[width=0.4\textwidth]{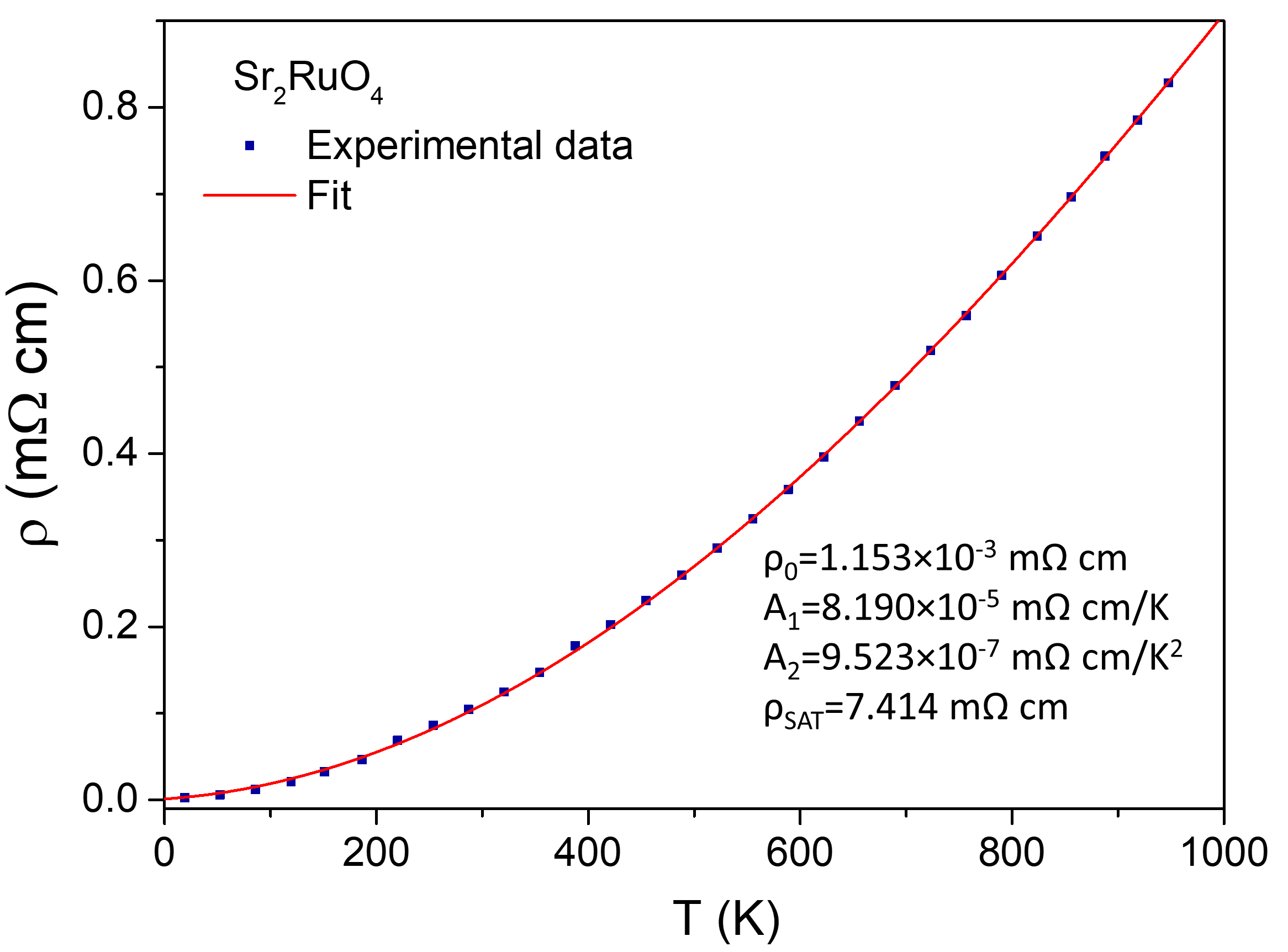}
\caption{Fitting of $\rho$(T) of Sr\textsubscript{2}RuO\textsubscript{4}. Data from Ref.\cite{tyler1998high}.}
\label{fig:metallic resistivity}
\end{figure}

\item \textbf{CeRu\textsubscript{2}Si\textsubscript{2}}

CeRu\textsubscript{2}Si\textsubscript{2} is well known as a canonical heavy fermion compound. This material is characterized by its rather large value of specific heat at low-$T$ and  a metamagnetic-like transition \cite{flouquet1995heavy}. The resistivity data of CeRu\textsubscript{2}Si\textsubscript{2} analysed here were obtained from the work of Besnus et al. \cite{besnus1985low}.

\renewcommand{\thefigure}{S17}
\begin{figure}[htp!]
\centering
\includegraphics[width=0.4\textwidth]{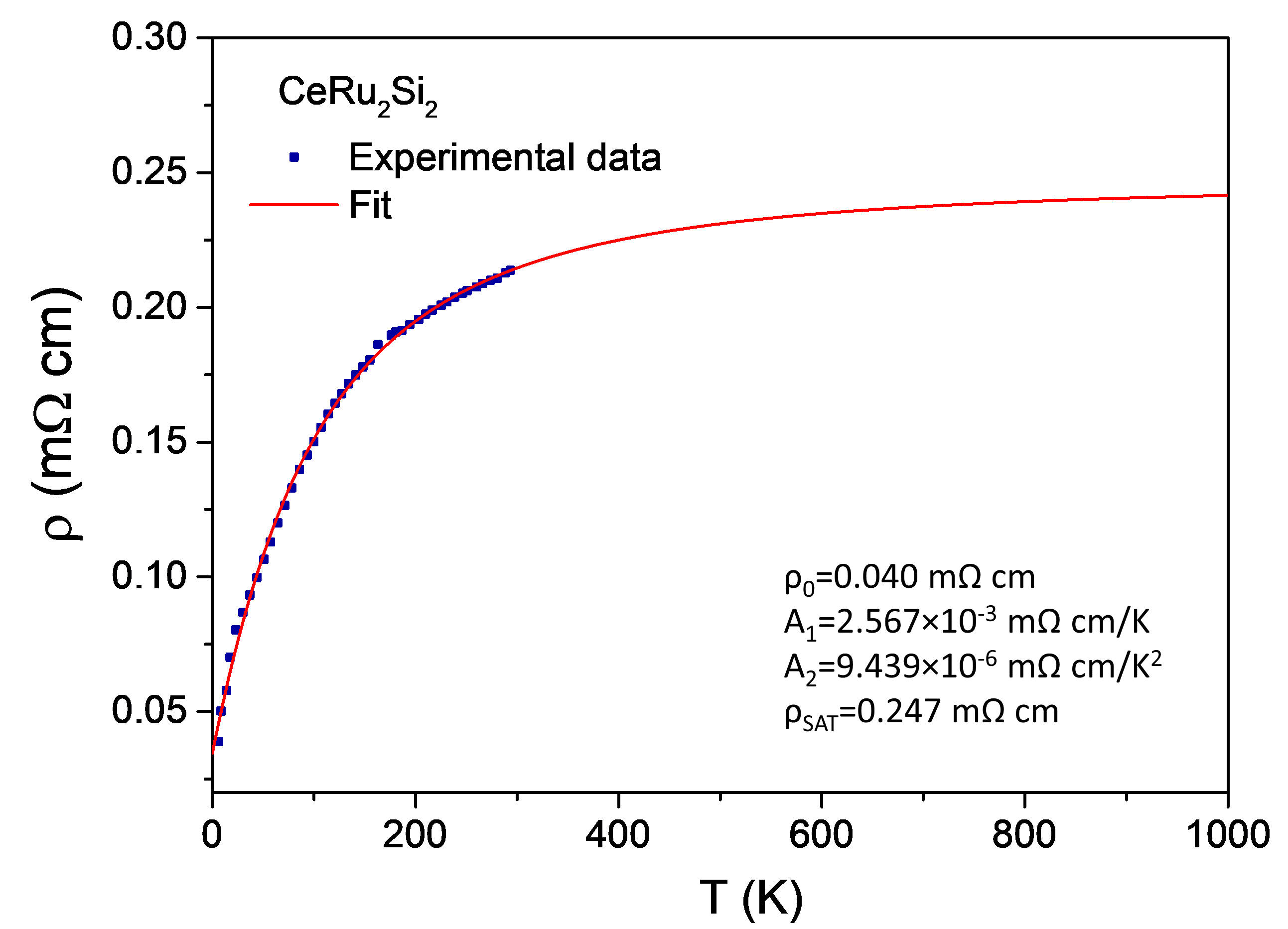}
\caption{Fitting of $\rho$(T) of CeRu\textsubscript{2}Si\textsubscript{2}. Data from Ref.\cite{besnus1985low}.}
\label{fig:metallic resistivity}
\end{figure}

\item \textbf{CrO\textsubscript{2}}

The resistivity data of CrO\textsubscript{2} were extracted from Ref. \cite{lewis1997band}. CrO\textsubscript{2} has also been discussed as bad metal. It shows signs of saturation but at higher values than that predicted by the MIR limit. The DC-PRF fit for this material is significantly worse than for other systems. We have also found that different sources in the literature show different behaviour \cite{suzuki1998resistivity,ranno1997production,hwang1997enhanced} so the intrinsic temperature dependence of resisitivity in this material still needs to be determined. 

\renewcommand{\thefigure}{S18}
\begin{figure}[htp!]
\centering
\includegraphics[width=0.4\textwidth]{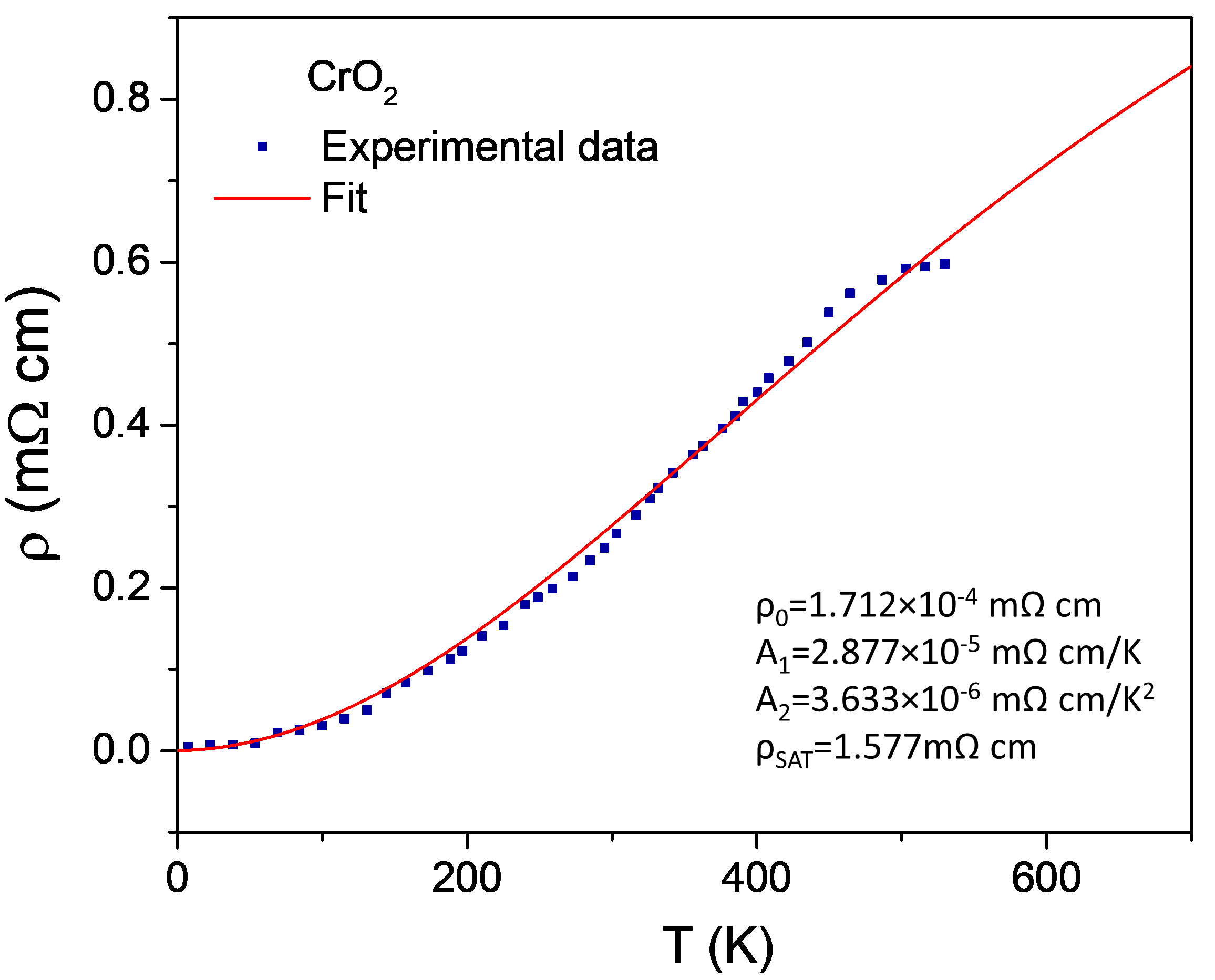}
\caption{Fitting of $\rho$(T) of CrO\textsubscript{2}. Data from Ref.\cite{lewis1997band}.}
\label{fig:metallic resistivity}
\end{figure}

\item \textbf{VO\textsubscript{2}}

 An extended measurement of resistivity up to 840 K was performed in VO\textsubscript{2} by Philip et al. \cite{allen1993resistivity}. Above the metal-insulator transition temperature ($\sim$ 350 K), the temperature dependence of resistivity in the metallic phase of VO\textsubscript{2} is linear. The calculated mean-free-path at 800 K, according to \cite{allen1993resistivity}, is only 3.3 \AA, which prompted these authors to propose unconventional behaviour.

\renewcommand{\thefigure}{S19}
\begin{figure}[htp!]
\centering
\includegraphics[width=0.4\textwidth]{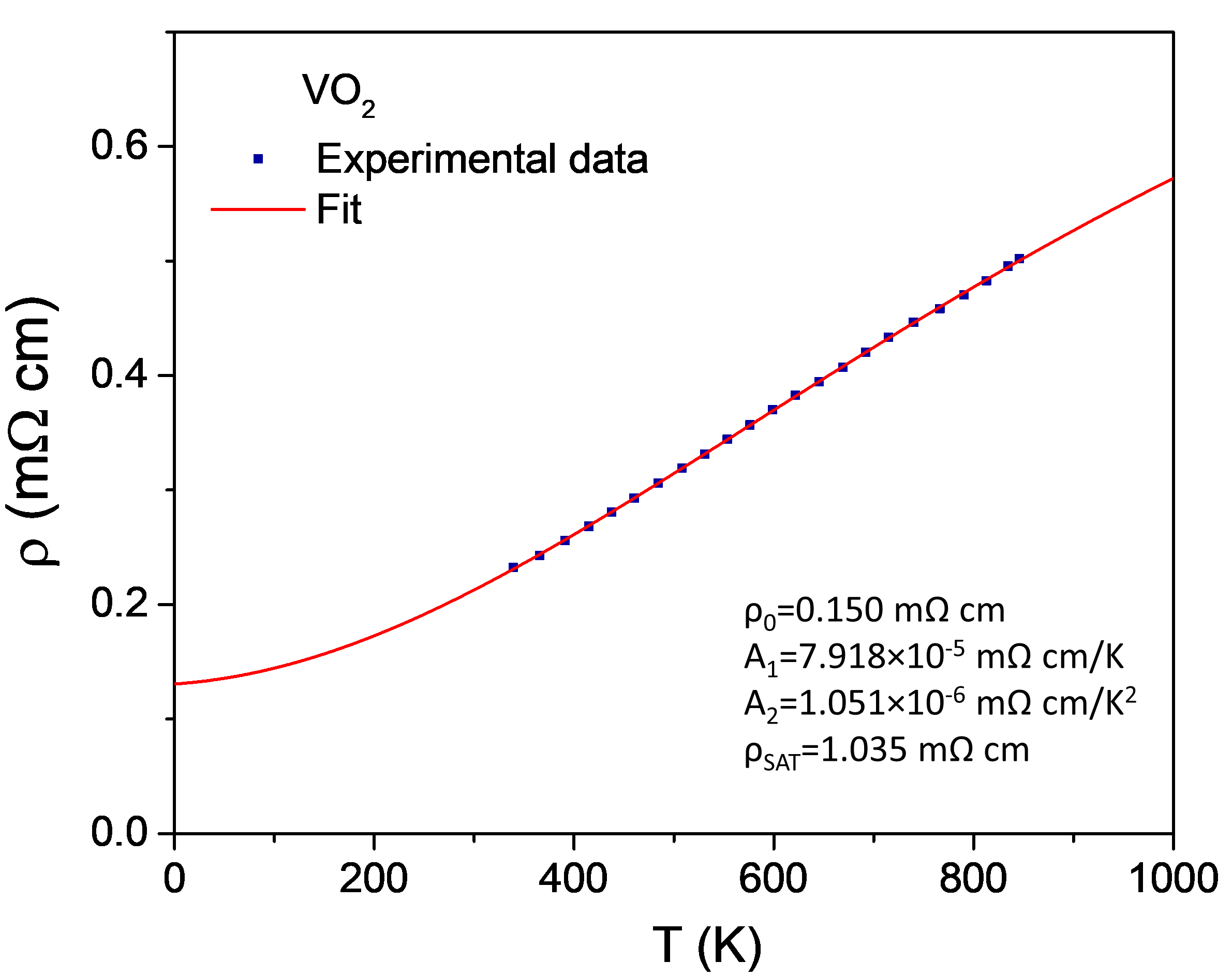}
\caption{Fitting of $\rho$(T) of VO\textsubscript{2}. Data from Ref.\cite{allen1993resistivity}.}
\label{fig:metallic resistivity}
\end{figure}

\item \textbf{Nb\textsubscript{3}Sb}

The A15 intermetallics have attracted particular interests because of their high superconducting transition temperatures. On the other hand, the obvious saturation of resistivity at temperature range below 1000 K, which is usually absent in most of metallic systems, make them ideal candidates for demonstrating the validity of Mott-Ioffe-Regel limit. Herein, Nb\textsubscript{3}Sb was chosen as an example and the data were extracted from Ref. \cite{fisk1976saturation}. Our fit to the experimental data gives rise to a value of $\rho$\textsubscript{sat}= 0.136 m$\Omega$ cm, very consistent with that of the predicted Mott-Ioffe-Regel limit (0.135-0.150 m$\Omega$ cm) in this family of compounds \cite{fisk1976saturation}.

\renewcommand{\thefigure}{S20}
\begin{figure}[htp!]
\centering
\includegraphics[width=0.4\textwidth]{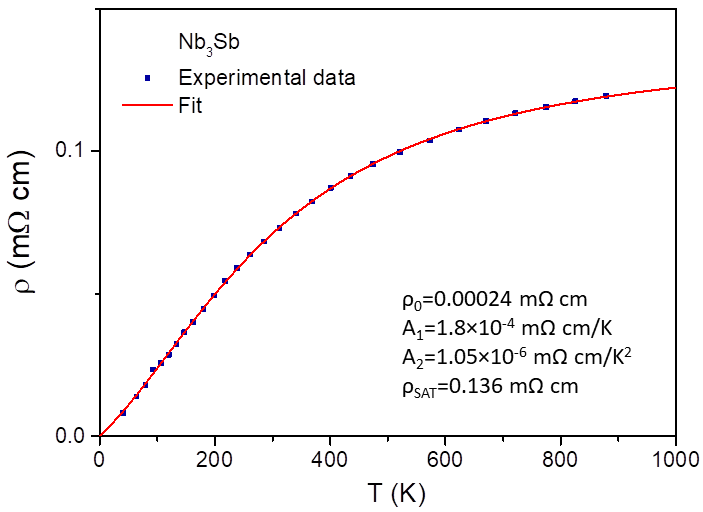}
\caption{Fitting of $\rho$(T) of Nb\textsubscript{3}Sb. Data from Ref.\cite{fisk1976saturation}.}
\label{fig:metallic resistivity}
\end{figure}

\item \textbf{SmNiO\textsubscript{3}}

In comparison with NdNiO\textsubscript{3} studied in this work, SmNiO\textsubscript{3} displays a higher metal-insulator transition temperature, above 400 K. An extended measurement of metallic resistivity has been reported by Jaramillo et al.\cite{jaramillo2014origins}. Moreover, they revealed a bad-metallic behaviour of this material on the basis of electrical and optical conductivity measurements. Indeed, the fit to the resistivity data extracted from the same work gave a saturation resistivity of $\sim$ 1.077 $\mu \Omega$ cm, which is obviously larger than the predicted MIR limit (0.5 $\mu \Omega$ cm) of this material. 

\renewcommand{\thefigure}{S21}
\begin{figure}[htp!]
\centering
\includegraphics[width=0.4\textwidth]{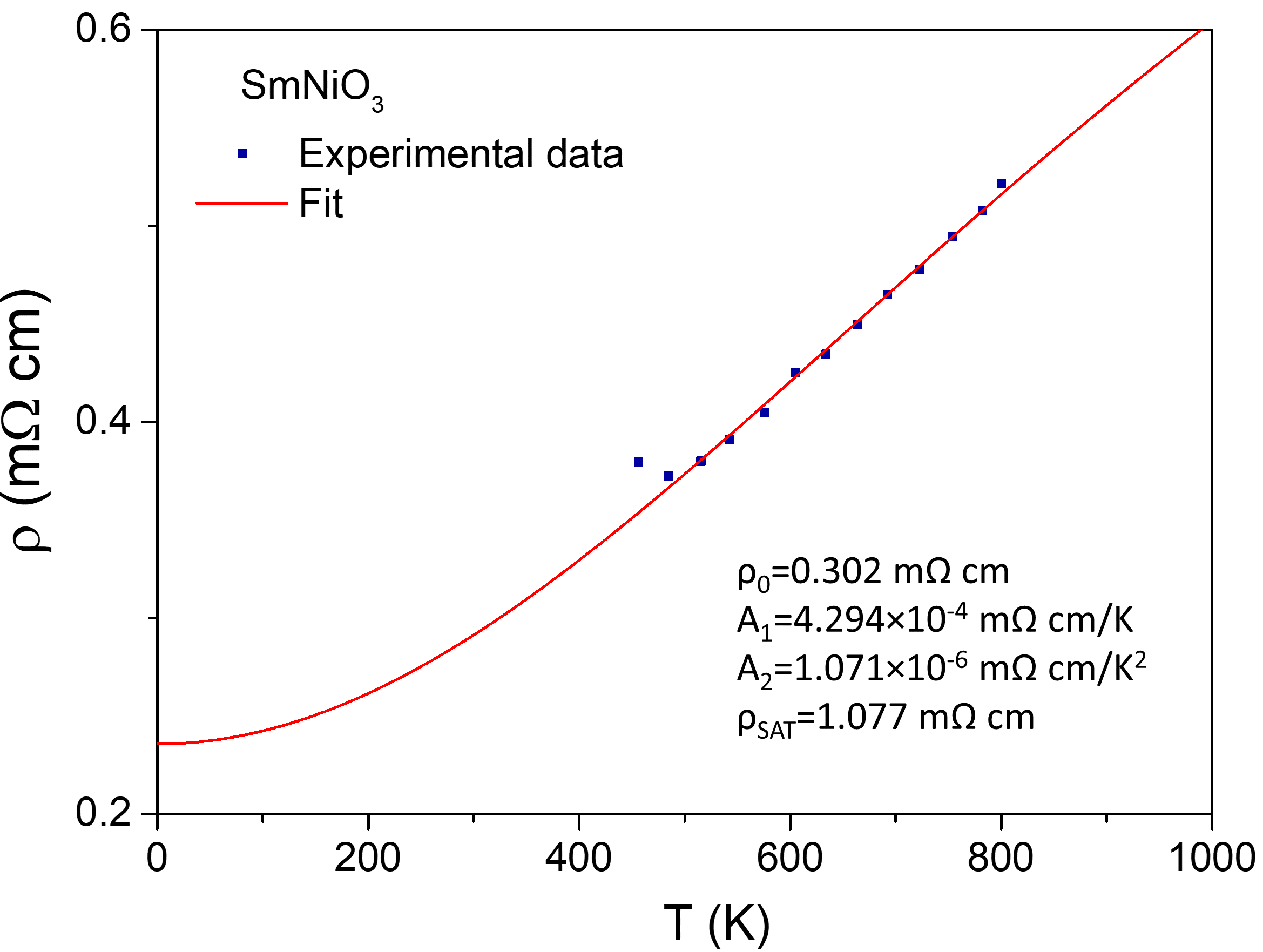}
\caption{Fitting of $\rho$(T) of SmNiO\textsubscript{3}. Data from Ref.\cite{jaramillo2014origins}.}
\label{fig:metallic resistivity}
\end{figure}

\item \textbf{Nd\textsubscript{0.825}Sr\textsubscript{0.175}NiO\textsubscript{2}}

Rare-earth nickelates have attracted renewed interests since the discovery of superconductivity in the related infinite-layer compound (Nd\textsubscript{1-x}Sr\textsubscript{x}NiO\textsubscript{2}) by Li et al. \cite{li2019superconductivity}. The data studied in the present work are extracted from the subsequent work from the same authors \cite{li2020superconducting}. In this work, Li et al. reported the phase diagram of Nd\textsubscript{1-x}Sr\textsubscript{x}NiO\textsubscript{2} infinite layer thin films grown on SrTiO\textsubscript{3}. In our present work, resistivity of Nd\textsubscript{0.825}Sr\textsubscript{0.175}NiO\textsubscript{2} was studied as it displays the best superconductivity and lowest resistivity among different doping levels.    

\renewcommand{\thefigure}{S22}
\begin{figure}[htp!]
\centering
\includegraphics[width=0.4\textwidth]{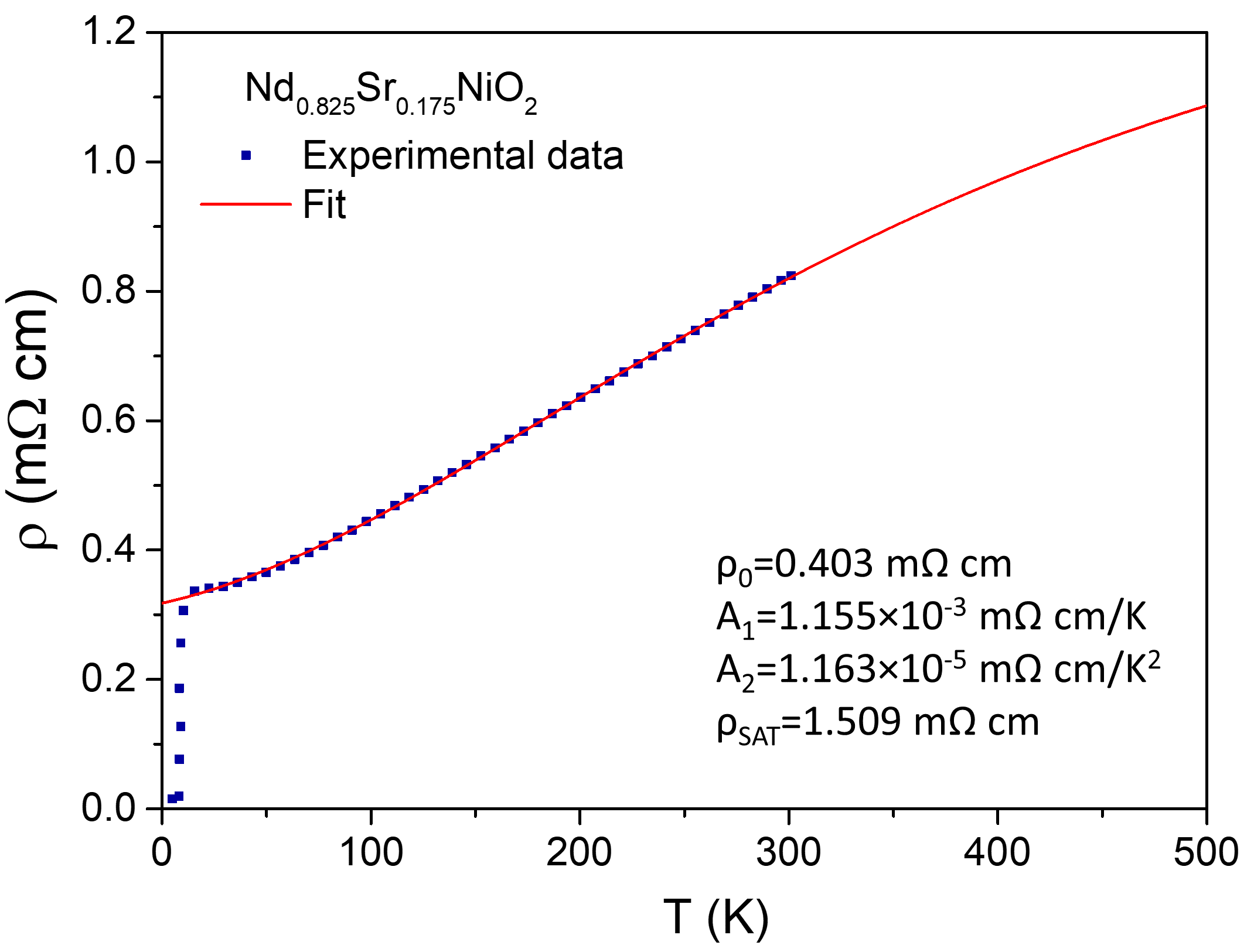}
\caption{Fitting of $\rho$(T) of Nd\textsubscript{0.825}Sr\textsubscript{0.175}NiO\textsubscript{2}. Data from Ref.\cite{li2020superconducting}.}
\label{fig:metallic resistivity}
\end{figure}

\end{itemize}

\subsection{Planckian metals}

\begin{itemize}

\item \textbf{Nd-LSCO and Bi2212}

The linear-$T$ resistivity in high-$T_c$ superconductors with optimized doping has been a major puzzle in condensed matter physics. In a systematically study by Legros et al. \cite{legros2019universal}, they revealed that the origin of the $T$-linear resistivity in many different cuprates is associated with a universal Planckian dissipation. Herein, the resistivity data of two systems: Nd-doped La\textsubscript{2-x}Sr\textsubscript{x}CuO\textsubscript{4} (Nd-LSCO) with p=0.24 and Bi\textsubscript{2}Sr\textsubscript{2}CaCu\textsubscript{2}O\textsubscript{8+$\delta$} (Bi2212) with p=0.23, measured under high magnetic field were also extracted and fit with the parallel resistor model. By applying a high magnetic field, the superconductivity transition is suppressed and the $T$-linear resistivity towards absolute zero kelvin is obtained. Notably, the comparable parameters obtained from our fit in these two systems manifest a similar origin of $T$-linear resistivity in different cuprates, which is well consistent with the conclusion of Legros et al.

\renewcommand{\thefigure}{S23}
\begin{figure}[htp!]
\centering
\includegraphics[width=0.4\textwidth]{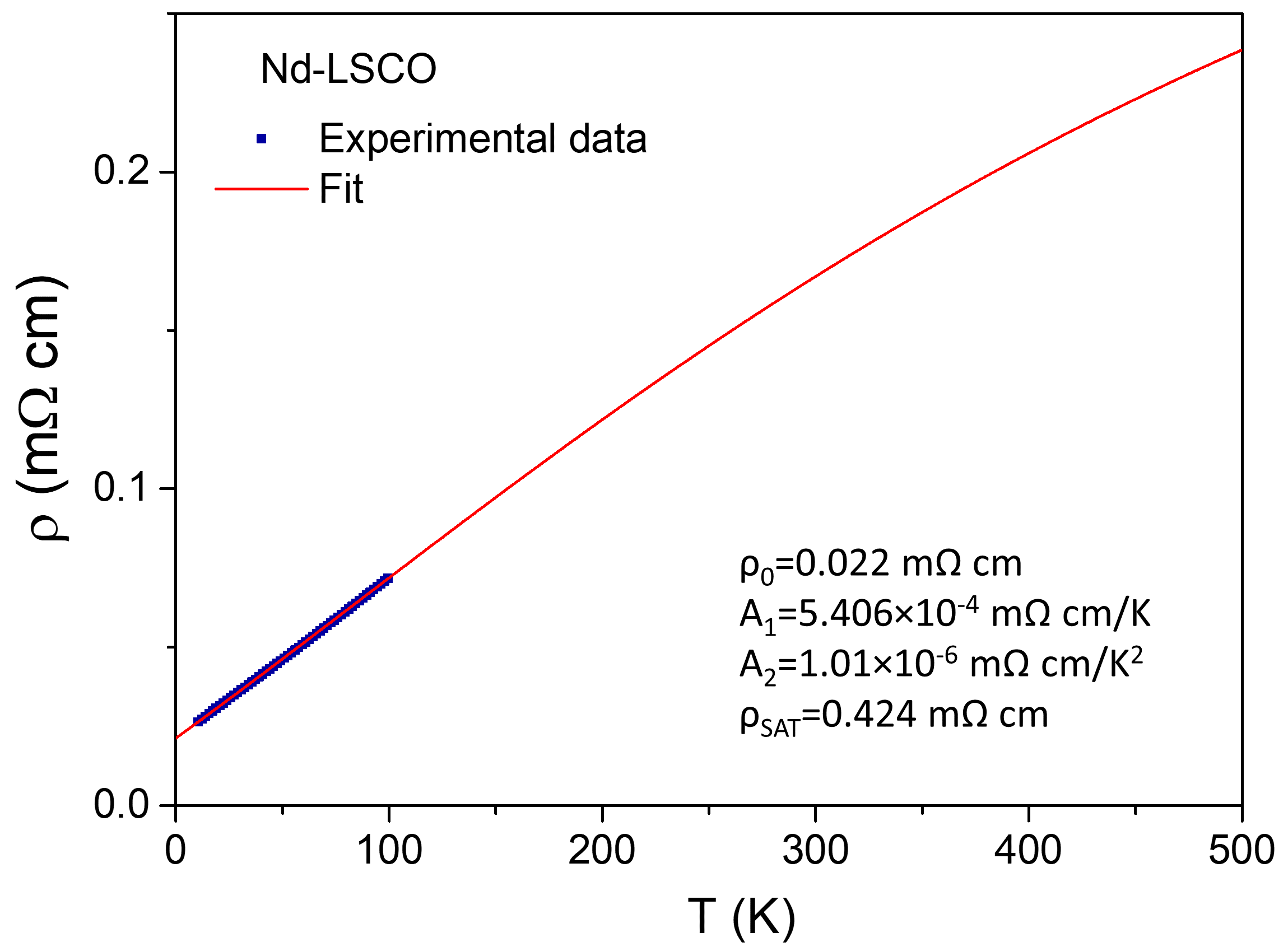}
\caption{Fitting of $\rho$(T) of Nd-doped La\textsubscript{2-x}Sr\textsubscript{x}CuO\textsubscript{4} with p=0.24 (H=16 T). Data from Ref.\cite{legros2019universal}.}
\label{fig:metallic resistivity}
\end{figure}

\renewcommand{\thefigure}{S24}
\begin{figure}[htp!]
\centering
\includegraphics[width=0.4\textwidth]{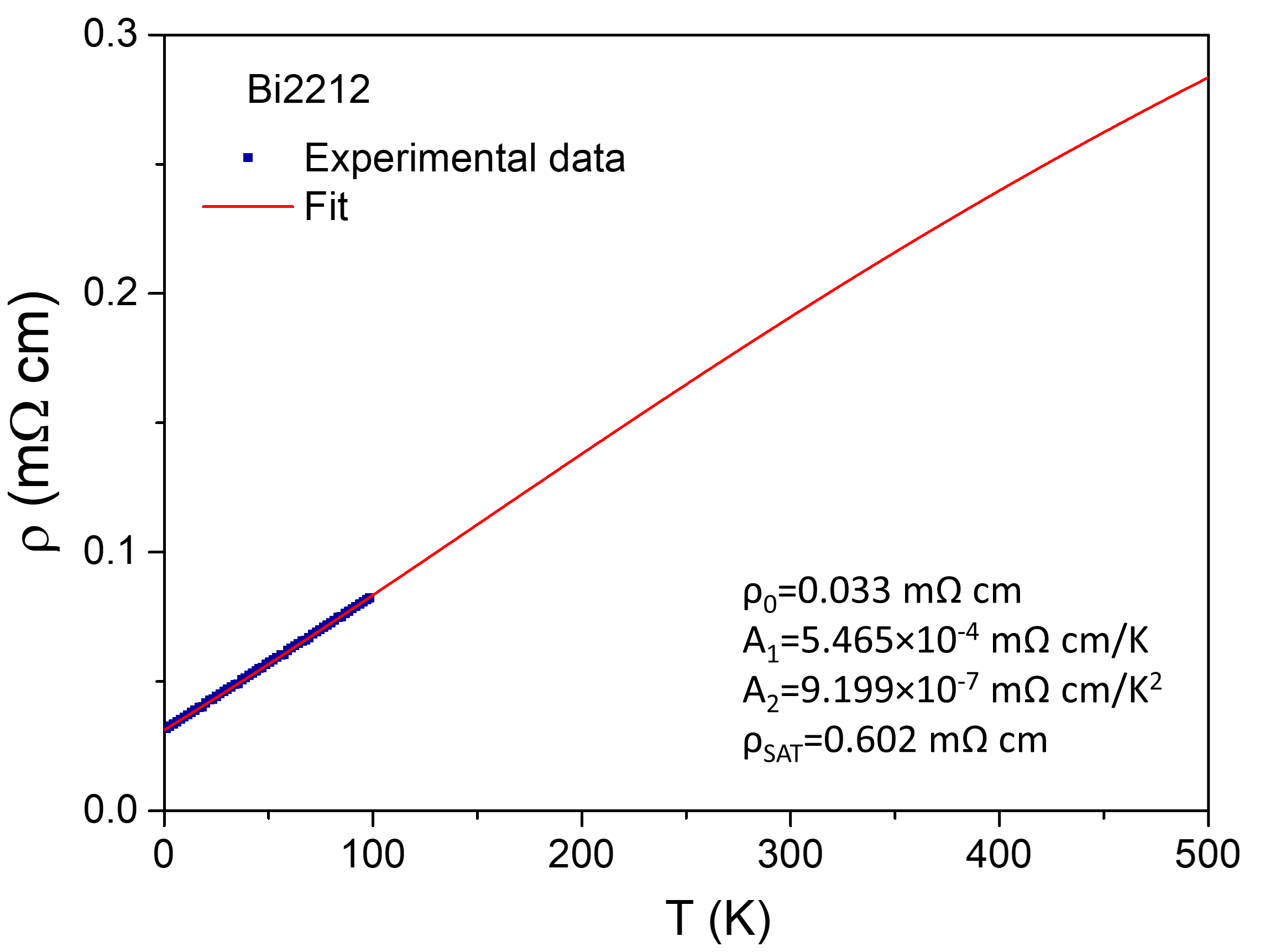}
\caption{Fitting of $\rho$(T) of Bi\textsubscript{2}Sr\textsubscript{2}CaCu\textsubscript{2}O\textsubscript{8+$\delta$} (Bi2212) with p=0.23 (H= 55 T). Data from Ref.\cite{legros2019universal}.}
\label{fig:metallic resistivity}
\end{figure}

\item \textbf{Sr\textsubscript{3}Ru\textsubscript{2}O\textsubscript{7}}

As a magnetic-field-tuned quantum system, Sr\textsubscript{3}Ru\textsubscript{2}O\textsubscript{7} can approach the quantum critical point under a critical field and, therefore, displays a low-$T$ linear-dependence of resistivity \cite{bruin2013similarity}. Bruin et al.\cite{bruin2013similarity} demonstrated that the linear-$T$-resistivity of Sr\textsubscript{3}Ru\textsubscript{2}O\textsubscript{7} under such conditions also arises from the approach of the scattering rate to the Planckian limit.

\renewcommand{\thefigure}{S25}
\begin{figure}[htp!]
\centering
\includegraphics[width=0.4\textwidth]{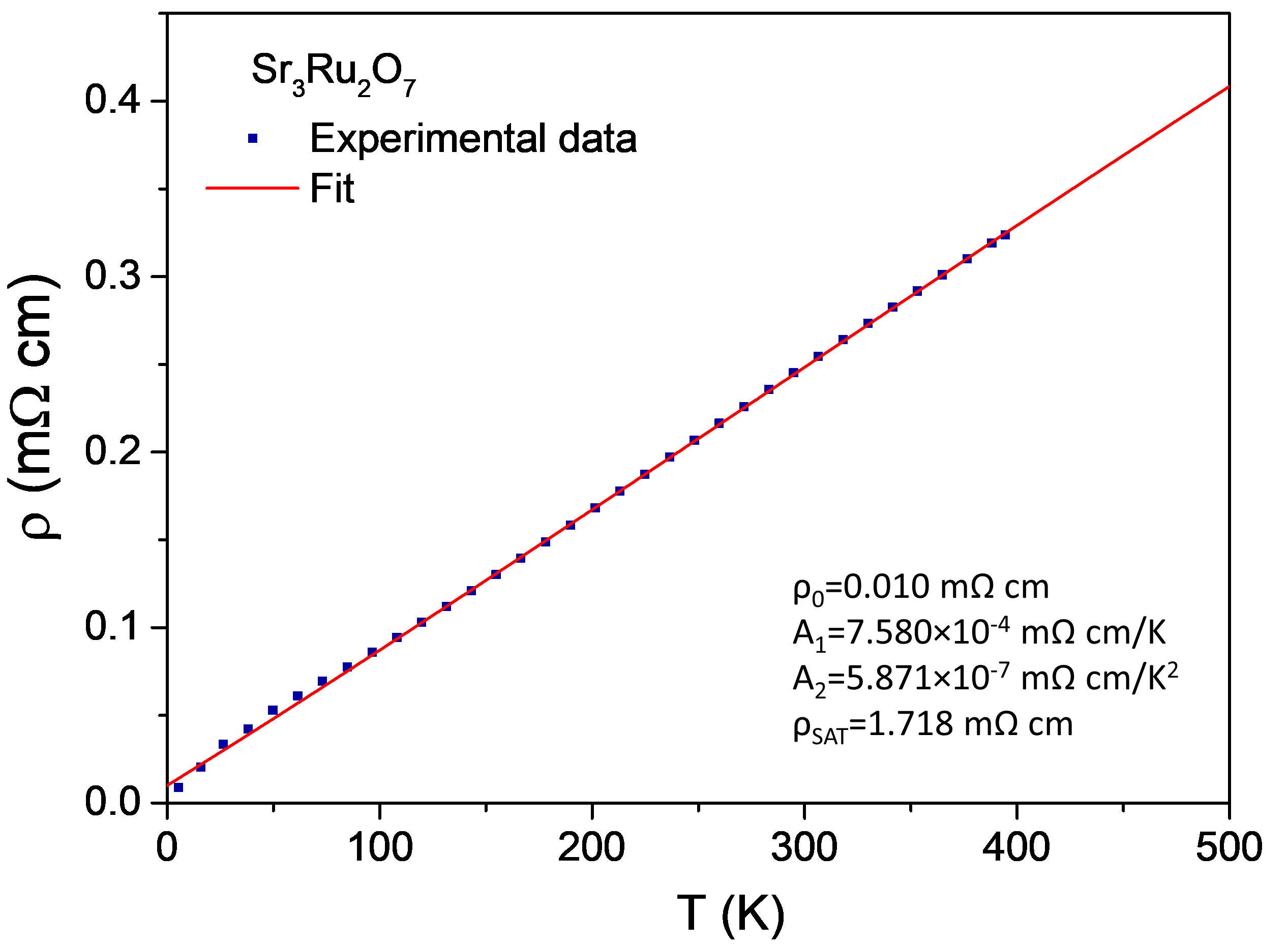}
\caption{Fitting of $\rho$(T) of Sr\textsubscript{3}Ru\textsubscript{2}O\textsubscript{7}. Data from Ref.\cite{bruin2013similarity}.}
\label{fig:metallic resistivity}
\end{figure}

\item \textbf{Ba(Fe\textsubscript{1/3}Co\textsubscript{1/3}Ni\textsubscript{1/3})\textsubscript{2}As\textsubscript{2}}

The resistivity data of Ba(Fe\textsubscript{1/3}Co\textsubscript{1/3}Ni\textsubscript{1/3})\textsubscript{2}As\textsubscript{2} were obtained from the work of ref. \cite{nakajima2020quantum}. The linear dependence of resistivity in this non-superconducting iron pnictide system has also been found to obey a universal scaling relation between temperature and applied magnetic fields down to the lowest energy scales\cite{nakajima2020quantum}.

\renewcommand{\thefigure}{S26}
\begin{figure}[htp!]
\centering
\includegraphics[width=0.4\textwidth]{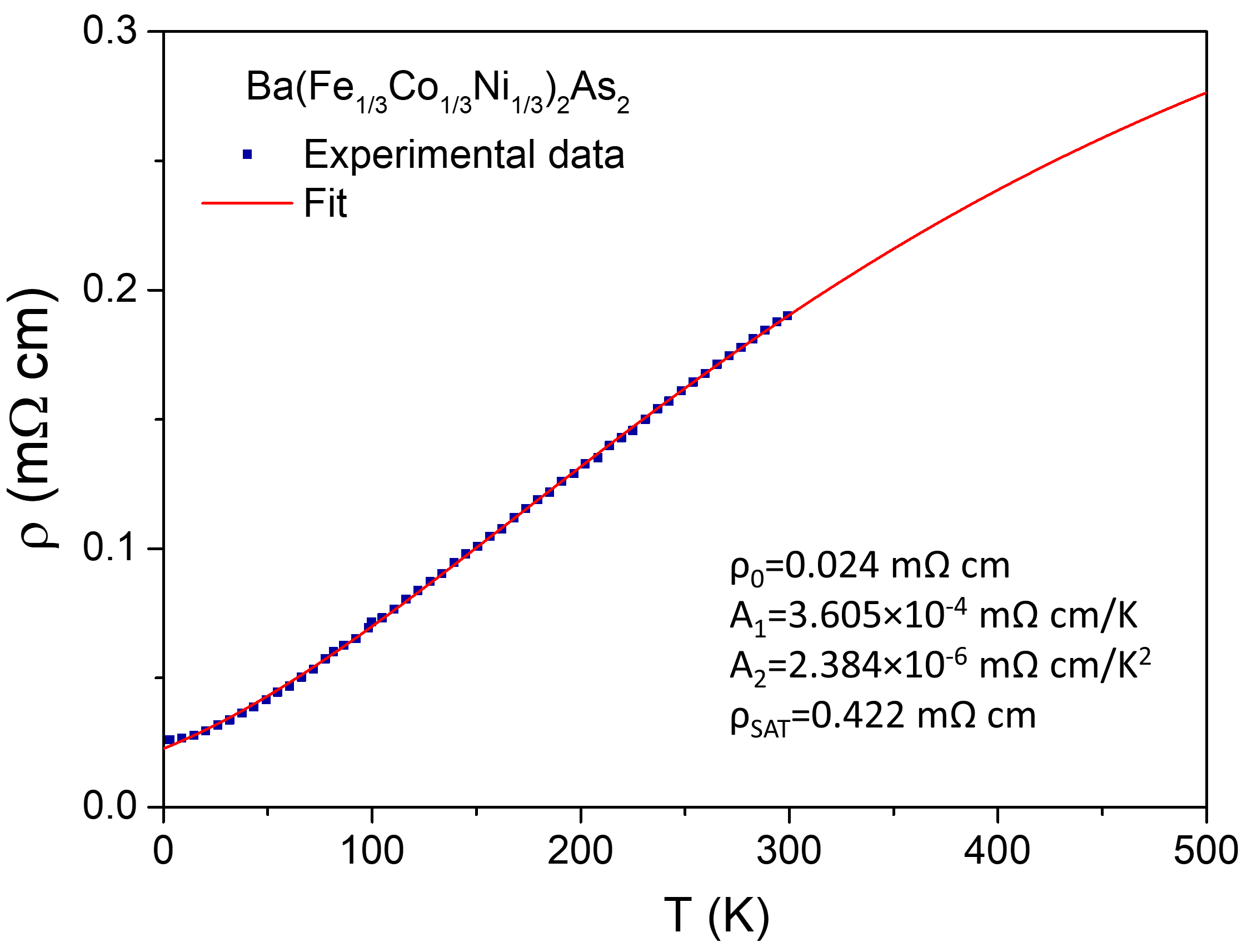}
\caption{Fitting of $\rho$(T) of Ba(Fe\textsubscript{1/3}Co\textsubscript{1/3}Ni\textsubscript{1/3})\textsubscript{2}As\textsubscript{2}. Data from Ref.\cite{nakajima2020quantum}.}
\label{fig:BFCNA}
\end{figure}

\item \textbf{NdNiO\textsubscript{3}}

\renewcommand{\thefigure}{S27}
\begin{figure}[htp!]
\centering
\includegraphics[width=0.4\textwidth]{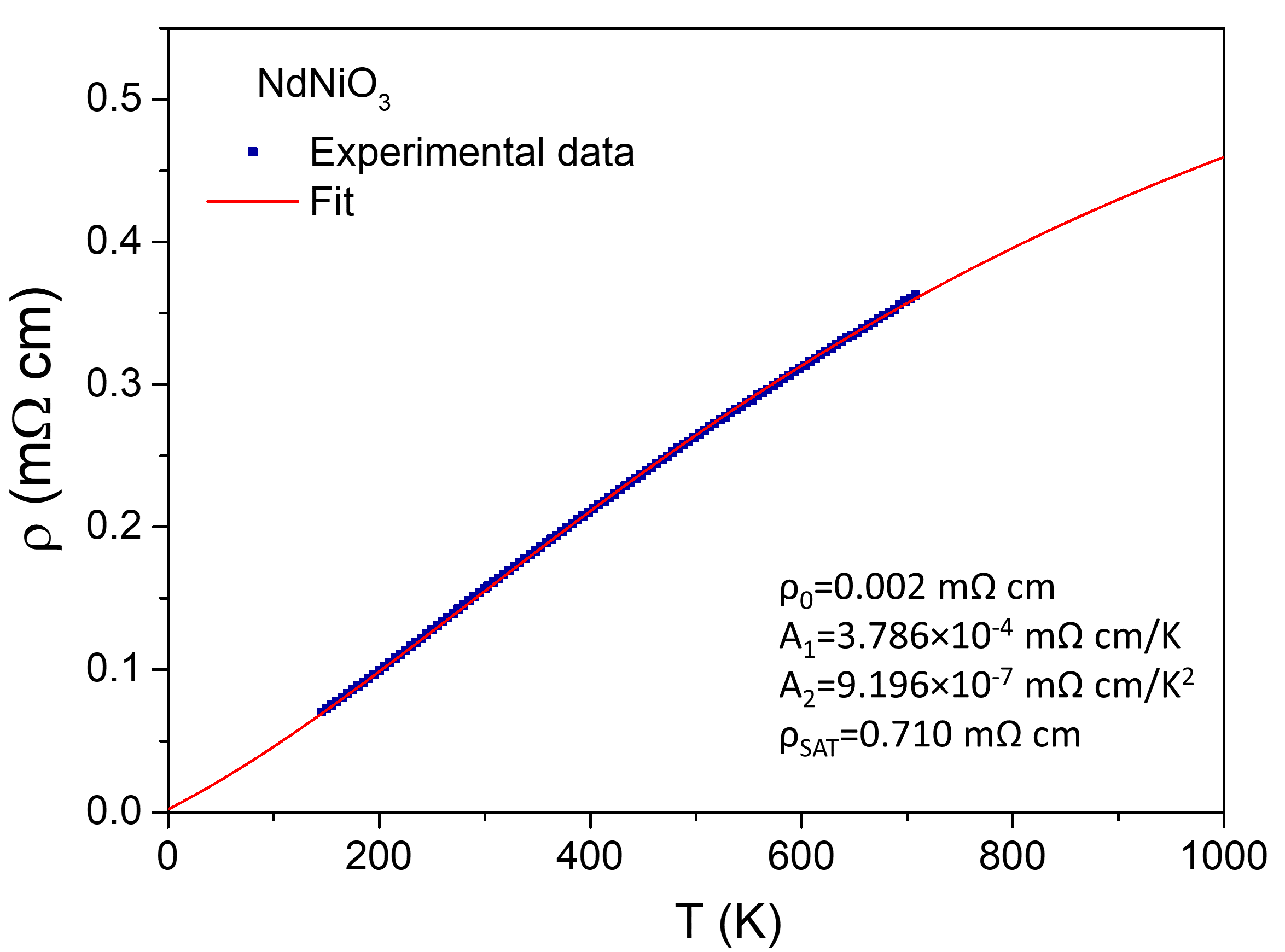}
\caption{Fitting of $\rho$(T) of NdNiO\textsubscript{3}. Data from present work.}
\label{fig:NNO}
\end{figure} 

\end{itemize}

\subsection{Simple metals}

For comparison, the resistivity of several simple metals is also included in this work. These simple metals had also been discussed as Planckian metals by Bruin et al. \cite{bruin2013similarity} The resistivity data of Cu and Nb was extracted from the work of Gunnarsson et al. \cite{gunnarsson2003colloquium}; while those of Al, Co, and Pd were obtained from Ref. \cite{de1988temperature}. Except for Nb, the resistivity of simple metals shows a linear-$T$ dependence at high-$T$. At low-$T$, a Fermi-liquid-like $T^2$ dependence is observed. 

However, these typical performance of resistivity is absent in Nb. Among simple transition metals, Nb plays special role due to its high superconducting transition temperature. The resistivity of Nb shows a significant saturation at ultra-high temperature. The fit to the $\rho$(T) of Nb gave a saturation resistivity about 100 $\mu \Omega$ cm, which is well consistent with that calculated by Gunnarsson et al. \cite{gunnarsson2003colloquium}. Moreover, as shown in Fig. 3c in main text, the Nb shows a significantly larger values of $\alpha_1$ compared to $\alpha_2$. Interestingly, Nb has also been found to show a more obvious deviation from the Planckian limit in comparison with other simple metals \cite{bruin2013similarity}. We believe these anomalous behaviours of Nb are all attributed to their large electron-phonon effects \cite{crabtree1987anisotropy}.

\renewcommand{\thefigure}{S28}
\begin{figure}[htp!]
\centering
\includegraphics[width=0.4\textwidth]{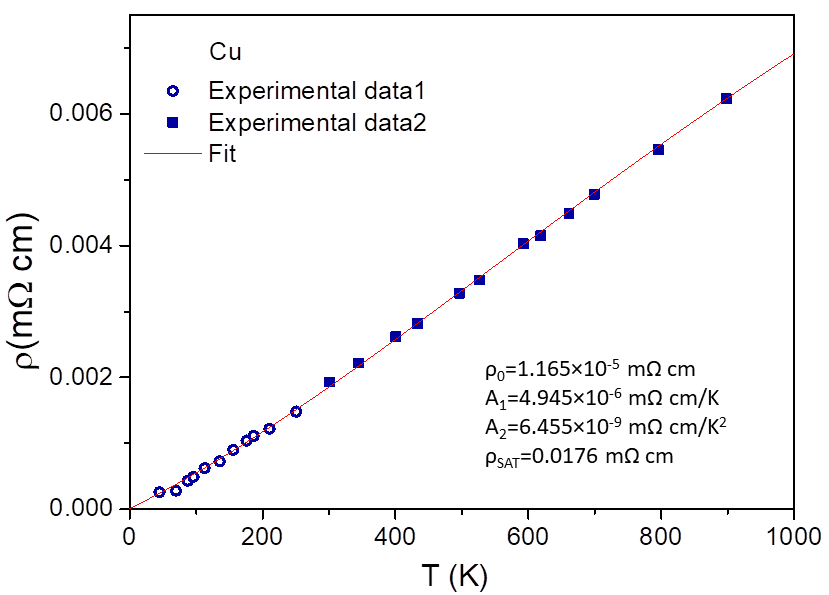}
\caption{Fitting of $\rho$(T) of Cu. Experimental data1 from Ref. \cite{allen1986dc} and Experimental data2 from Ref. \cite{gunnarsson2003colloquium}.}
\label{fig:Cu}
\end{figure}

\renewcommand{\thefigure}{S29}
\begin{figure}[htp!]
\centering
\includegraphics[width=0.4\textwidth]{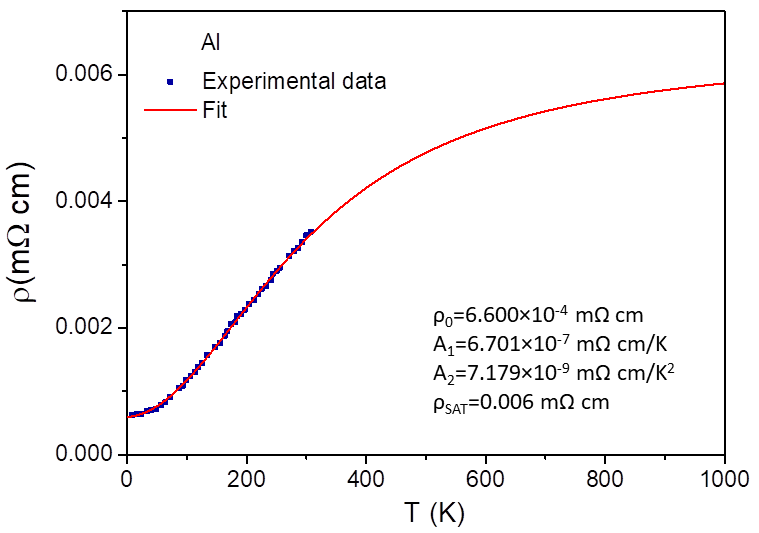}
\caption{Fitting of $\rho$(T) of Al. Data from Ref.\cite{de1988temperature}.}
\label{fig:Al}
\end{figure}

\renewcommand{\thefigure}{S30}
\begin{figure}[htp!]
\centering
\includegraphics[width=0.4\textwidth]{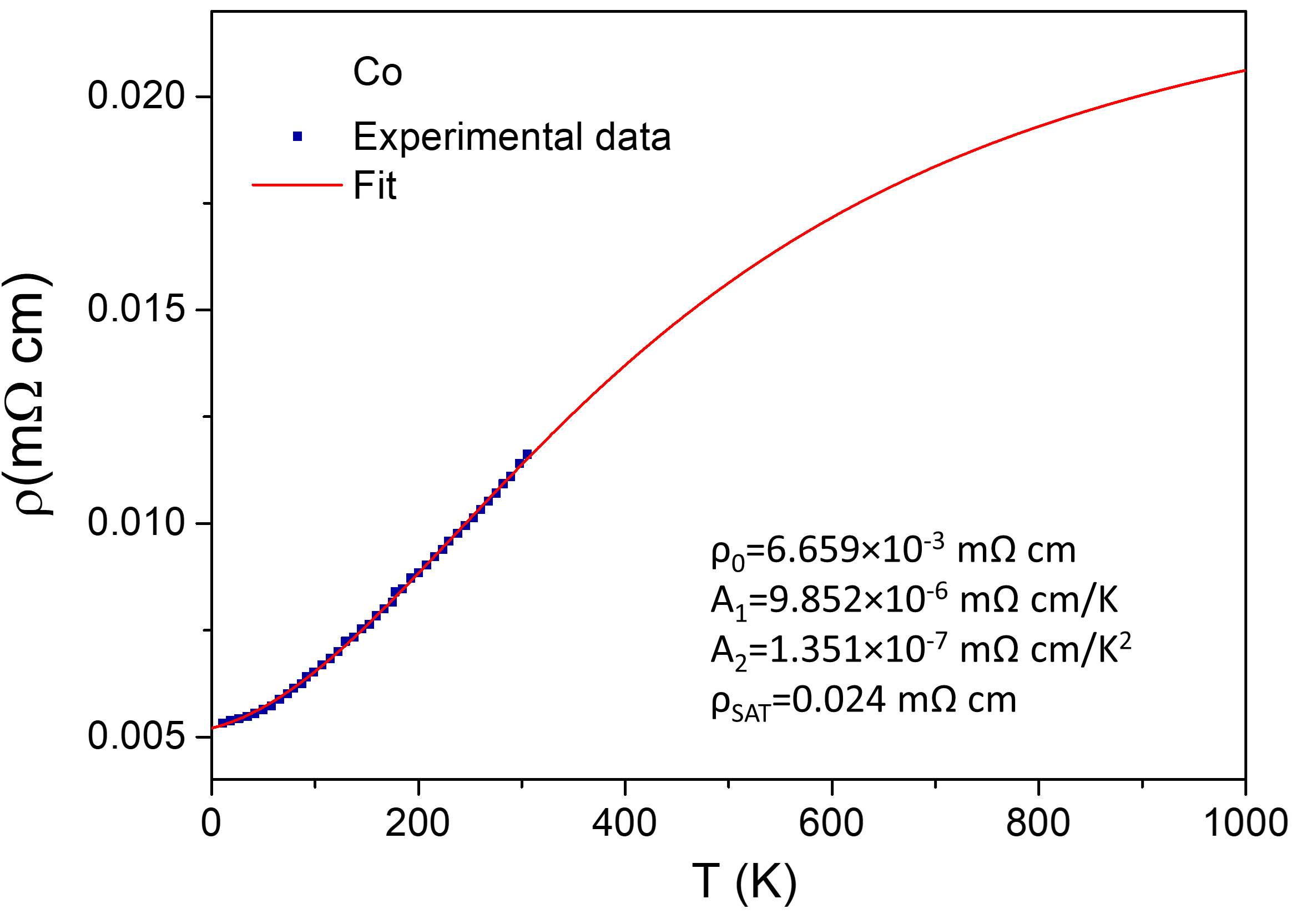}
\caption{Fitting of $\rho$(T) of Co. Data from Ref.\cite{de1988temperature}.}
\label{fig:Co}
\end{figure}

\renewcommand{\thefigure}{S31}
\begin{figure}[htp!]
\centering
\includegraphics[width=0.4\textwidth]{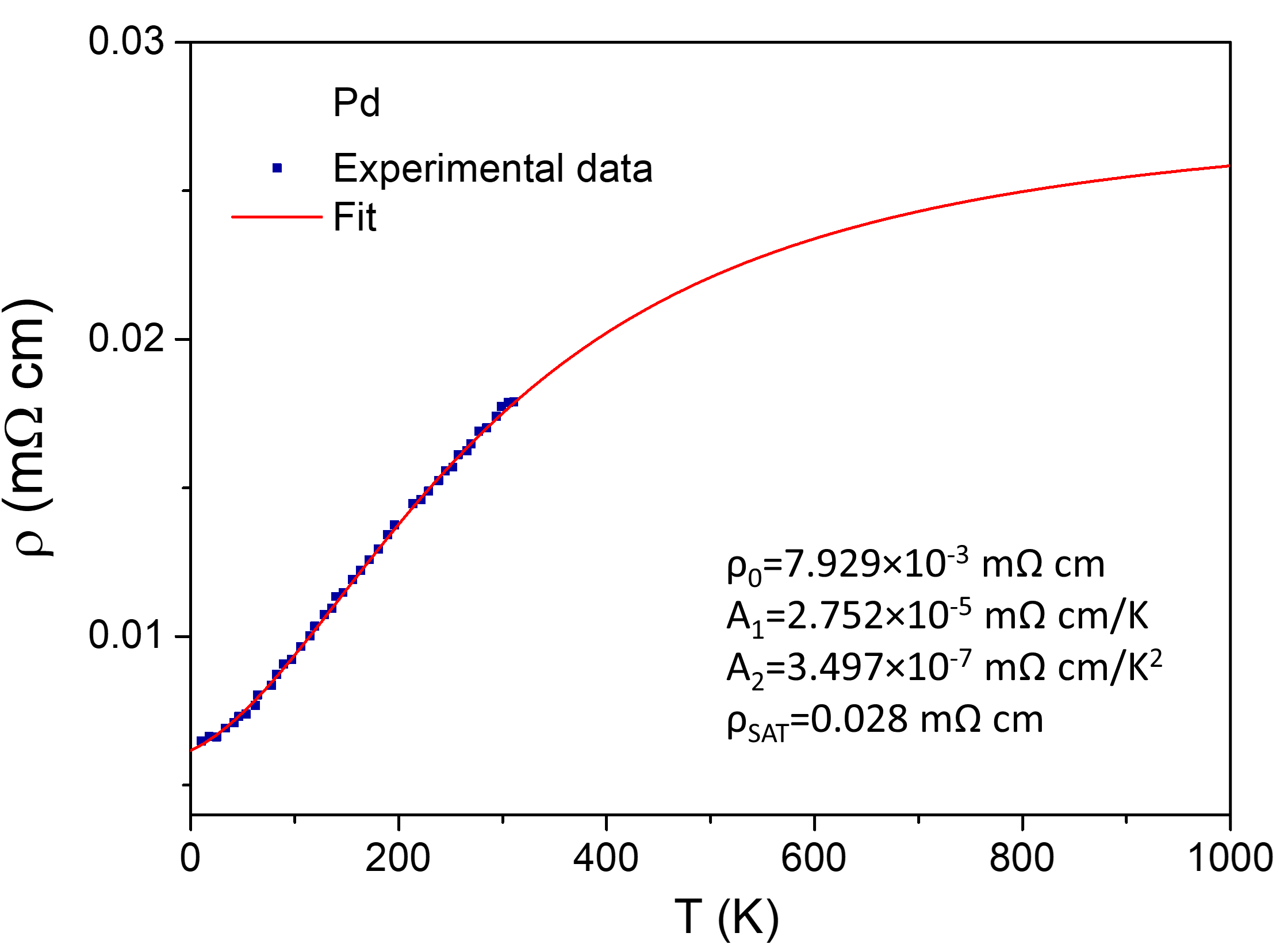}
\caption{Fitting of $\rho$(T) of Pd. Data from Ref.\cite{de1988temperature}.}
\label{fig:Pd}
\end{figure}

\renewcommand{\thefigure}{S32}
\begin{figure}[htp!]
\centering
\includegraphics[width=0.4\textwidth]{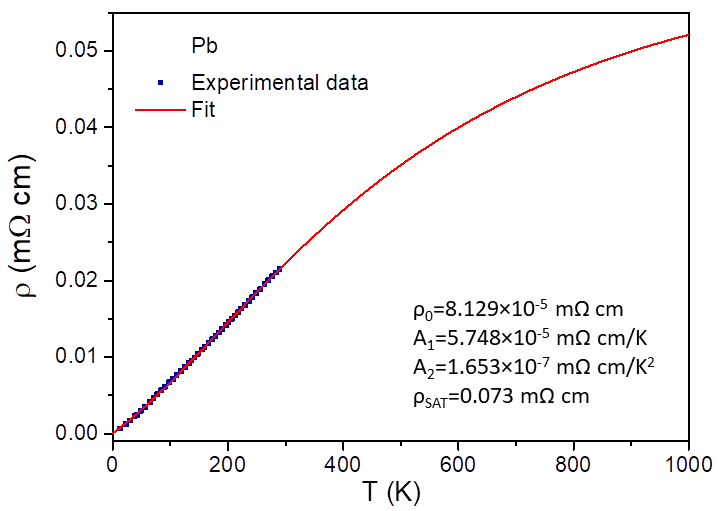}
\caption{Fitting of $\rho$(T) of Pb. Data from Ref.\cite{eiling1981pressure}.}
\label{fig:Pb}
\end{figure} 

\renewcommand{\thefigure}{S33}
\begin{figure}[htp!]
\centering
\includegraphics[width=0.4\textwidth]{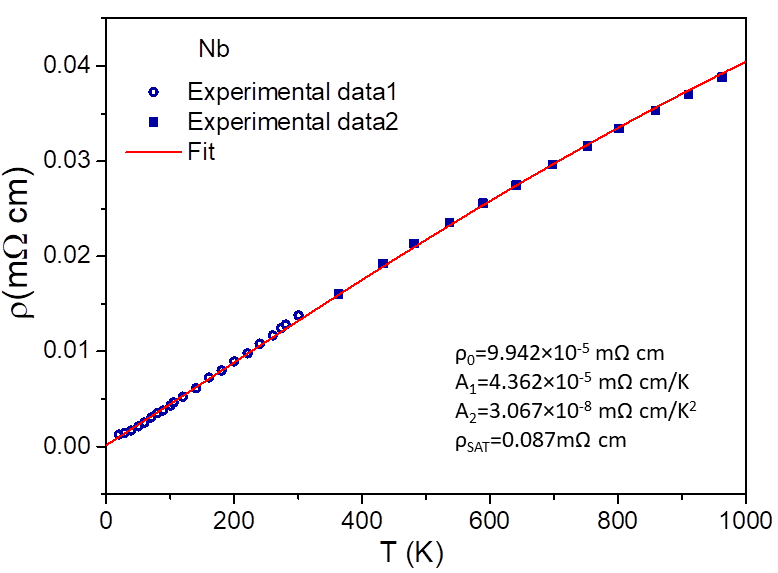}
\caption{Fitting of $\rho$(T) of Nb. Experimental data1 from Ref. \cite{allen1986dc} and Experimental data2 from Ref.\cite{gunnarsson2003colloquium}.}
\label{fig:Nb}
\end{figure}

\subsection{Effective mass and carrier density}

Values of effective mass (m*) and carrier density (n) used in Fig. 3c and discussed in the main text are obtained from refs. \cite{legros2019universal, varshney2006effect,padilla2005constant, jaramillo2014origins, ashcroft1976solid} and summarized in Table 1. The last column shows $A_1$n/m* in units of $k$\textsubscript{B}/$\hbar$ $e$\textsuperscript{2}. For those metals with simple isotropic Fermi surfaces, $A_1$n/m*= $k$\textsubscript{B}/$\hbar$ $e$\textsuperscript{2} at the Plackian dissipation limit \cite{legros2019universal}.

\renewcommand{\thetable}{S1}
\begin{table}[htp!]
\centering
\caption{\textbf{Relevant parameters for different reported metals.} Error bars for m* and n are obtained from the refs. \cite{legros2019universal, varshney2006effect,padilla2005constant, jaramillo2014origins, ashcroft1976solid}; while error bars in left two columns are calculated from the uncertainty of related parameters. }
\label{tab:my-table}
\resizebox{\columnwidth}{!}{
\begin{tabular}{|l|l|l|l|l|}
\hline
 Material & n(10\textsuperscript{27}m\textsuperscript{-3}) & m\textsuperscript{*}/m\textsubscript{0} & m*/n (m\textsubscript{0}/10\textsuperscript{27}m\textsuperscript{-3})  & $A_1$n/m* ($k_B/\hbar e^2$) \\ \hline

Bi2212 (p=0.23) & 6.8 & 8.4 $\pm$ 1.6 \cite{legros2019universal}& 1.2 $\pm$ 0.2& 1.2 $\pm$ 0.3\\\hline 
Nd-LSCO (p=0.24) & 7.9 & 12 $\pm$ 4 \cite{legros2019universal}& 1.5 $\pm$ 0.5 & 0.8 $\pm$ 0.3\\\hline 
LSCO (p=0.26) & 7.8 & 9.8 $\pm$ 1.7 \cite{legros2019universal}& 1.3 $\pm$ 0.2 &  0.7 $\pm$ 0.2\\\hline 
Rb\textsubscript{3}C\textsubscript{60} & 3.9 $\pm$ 0.5 & 3.6 $\pm$ 0.5\cite{varshney2006effect}& 0.9 $\pm$ 0.3 & 0.3 $\pm$ 0.2\\\hline
La\textsubscript{1.96}Sr\textsubscript{0.04}CuO\textsubscript{4} & 3 $\pm$ 0.5 & 4 $\pm$ 0.5 \cite{padilla2005constant}& 1.3 $\pm$ 0.4 & 0.7 $\pm$ 0.2\\\hline 
NdNiO\textsubscript{3} & 10 $\pm$ 8 & 7 $\pm$ 1 \cite{jaramillo2014origins}& 0.7 $\pm$ 0.1 & 1.32 $\pm$ 1.1\\\hline 
Cu & 85 & 1.3 \cite{ashcroft1976solid}& 0.015 & 0.27 $\pm$ 0.1\\\hline 
Nb & 52 & 12\cite{ashcroft1976solid} & 0.23 & 0.4 $\pm$ 0.2\\\hline 
Al & 60  & 1.4 \cite{ashcroft1976solid}& 0.023  & 0.6 $\pm$ 0.2\\\hline

 \end{tabular}
}

\end{table}

\subsection{Taylor expansion }

As shown in Fig. S34, in all cases $A$\textsubscript{2} $<$ $A$\textsubscript{1}. In most cases, $A$\textsubscript{2} $\ll$ $A$\textsubscript{1}, which justifies neglecting higher order terms. However, in some cases, such as Bi\textsubscript{2}Sr\textsubscript{2}Ca\textsubscript{0.89}Y\textsubscript{0.11}Cu\textsubscript{2}O\textsubscript{y},
Rb\textsubscript{3}C\textsubscript{60},
and YBa\textsubscript{2}Cu\textsubscript{3}O\textsubscript{6.45},
the value of $A$\textsubscript{2} is nearly of the same magnitude as that of $A$\textsubscript{1}, manifesting that more terms may be necessary in the expression of $\rho$\textsubscript{ideal}. Indeed, as shown in Fig. S13 to S15, the fit to the experimental data of these three systems is not as good as others. In these cases, we added the $T$\textsuperscript{3} term into the formalism of $\rho$\textsubscript{ideal}: 

\begin{equation}\label{definiation of ideal resistivity}
\rho \textsubscript{ideal}(T)=\rho_{0}+A_1 T + A_2 T^2+ A_3 T^3
\end{equation}
In the fit, the coefficient $A$\textsubscript{1} of the $T$-linear term was fixed, while the other three parameters are free. Fig. S35 to Fig. S37 show that the fits with the Eq. S(1) give rise to a better description of the experimental data compared to that shown in Section 2.2.

\renewcommand{\thefigure}{S34}
\begin{figure}[htp!]
\centering
\includegraphics[width=0.5\textwidth]{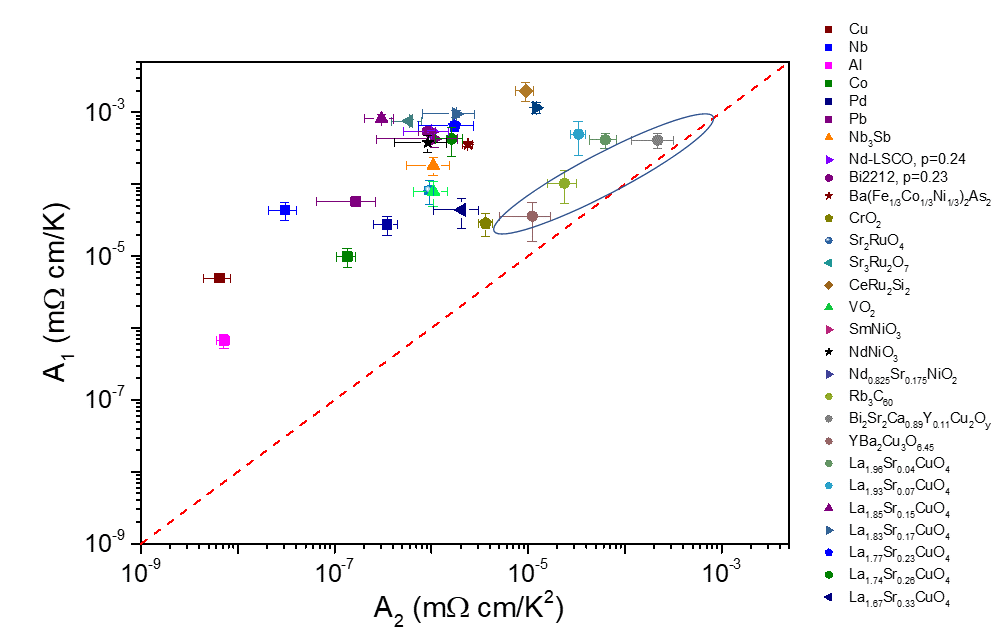}
\caption{$A$\textsubscript{1} \textit{versus} $A$\textsubscript{2}. The red dash line denotes $A$\textsubscript{1}= $A$\textsubscript{2}. The blue circle highlights three systems for which $A$\textsubscript{2} has nearly the same magnitude as that of $A$\textsubscript{1}. }
\label{fig:Nb}
\end{figure}

\renewcommand{\thefigure}{S35}
\begin{figure}[htp!]
\centering
\includegraphics[width=0.4\textwidth]{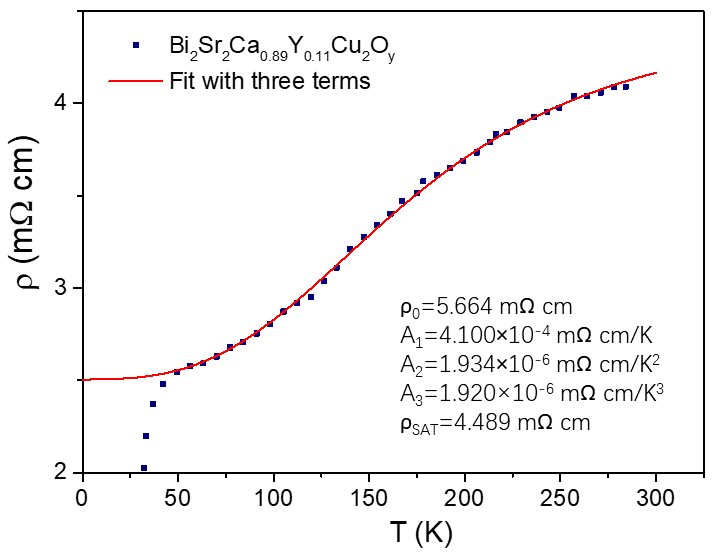}
\caption{New fit to Bi\textsubscript{2}Sr\textsubscript{2}Ca\textsubscript{0.89}Y\textsubscript{0.11}Cu\textsubscript{2}O\textsubscript{y} with Eq. S(1). The value of $A$\textsubscript{1} has been fixed to that obtained in Fig. S13.}
\label{fig:Nb}
\end{figure} 

\renewcommand{\thefigure}{S36}
\begin{figure}[htp!]
\centering
\includegraphics[width=0.4\textwidth]{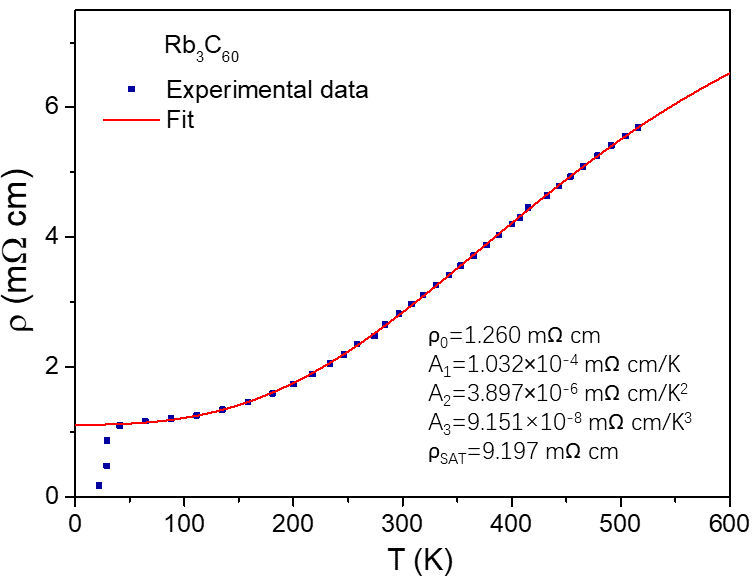}
\caption{New fit to Rb\textsubscript{3}C\textsubscript{60} with Eq. S(1). The value of $A$\textsubscript{1} has been fixed to that obtained in Fig. S14.}
\label{fig:Nb}
\end{figure} 

\renewcommand{\thefigure}{S37}
\begin{figure}[htp!]
\centering
\includegraphics[width=0.4\textwidth]{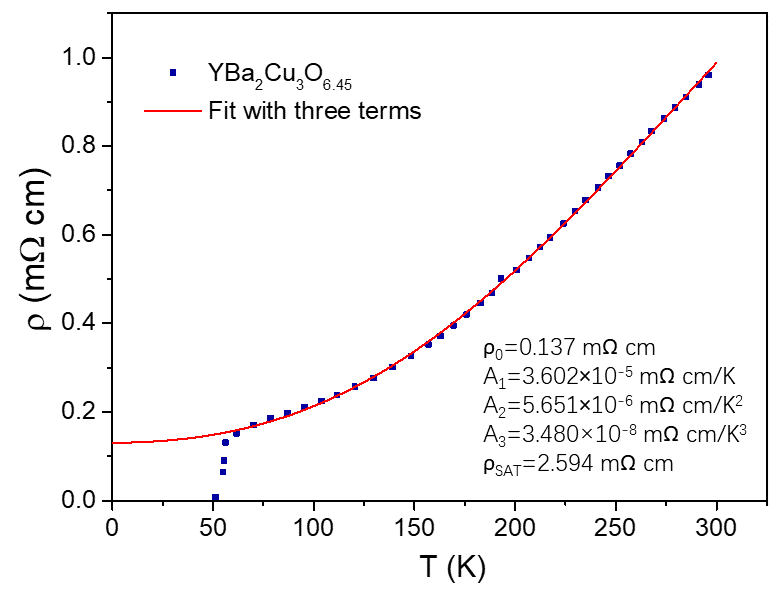}
\caption{New fit to YBa\textsubscript{2}Cu\textsubscript{3}O\textsubscript{6.45}. The value of $A$\textsubscript{1} has been fixed to that obtained in Fig. S15.}
\label{fig:Nb}
\end{figure}

\subsection{Determination of errors}

Except for NdNiO\textsubscript{3}, all the resistivity data were extracted from the plots in literature by software. Therefore, the uncertainty of $\rho$\textsubscript{300K} was determined by the quality of the extraction. The other parameters, such as $A$\textsubscript{1}, $A$\textsubscript{2}, and $\rho$\textsubscript{SAT}, were obtained from the uncertainty of the fit. The errors of $m$*/$n$ of several systems were calculated from the uncertainties of $m$* and $n$, which were obtained from the literature. 

All the parameters of all the fits and their errors have been provided as an open data file. However, it is worth to emphasize the difficulty in determining the exact values of parameter errors as the source data of resistivity are not available in literature. In this case, the uncertainty has already existed in the resistivity itself.

\clearpage

\medskip

\bibliography{Reference.bib}

\end{document}